\documentclass[superscriptaddress,groupedaddress,nofootinbib,preprintnumbers,12pt]{revtex4-1} 
\usepackage[totalheight = 23cm, totalwidth = 17cm]{geometry}
\usepackage{amssymb,amsmath,amsfonts,amsbsy}
\usepackage{bm}
\usepackage{graphicx}  
\usepackage{dcolumn}   
\usepackage{bm}        
\usepackage{amssymb}   
\usepackage{amsmath}
\usepackage{mathtools}
\usepackage{cases}
\usepackage{mathrsfs}
\usepackage{xcolor}
\usepackage{enumitem}
\usepackage{subfigure}
\usepackage[misc]{ifsym}
\usepackage{hyperref}
\usepackage{color}
\makeatletter
\newcommand*\bigcdot{\mathpalette\bigcdot@{.5}}
\newcommand*\bigcdot@[2]{\mathbin{\vcenter{\hbox{\scalebox{#2}{$\m@th#1\bullet$}}}}}
\makeatother
\newcommand{\mpl}{M_{\text{Pl}}}
\newcommand{\calH}{{\cal H}}

\newcommand{\calP}{{\cal P}}
\newcommand{\calR}{{\cal R}}
\newcommand{\be}{\begin{equation}}
\newcommand{\ee}{\end{equation}}
\newcommand{\nn}{\nonumber}

\newcommand*{\dif}{\mathop{}\!\mathrm{d}}

\begin{document}
	\preprint{IPMU22-0047}
	\title{Primordial Black Hole Formation\\ in Starobinsky's Linear Potential Model}
	\author{Shi Pi${}^{a,b,c}$} \email{shi.pi@itp.ac.cn} 
	\author{Jianing Wang${}^{a,d}$}\email{wangjianing@itp.ac.cn}
		\affiliation{
		$^{a}$ CAS Key Laboratory of Theoretical Physics, Institute of Theoretical Physics, Chinese Academy of Sciences, Beijing 100190, China \\
  $^{b}$ Center for High Energy Physics, Peking University, Beijing 100871, China\\
		$^{c}$ Kavli Institute for the Physics and Mathematics of the Universe (WPI), Chiba 277-8583, Japan\\
		$^{d}$ School of Physical Sciences, University of Chinese Academy of Sciences, Beijing 100049, China}
	\date{\today}
	\begin{abstract}
	We study the power spectrum of the comoving curvature perturbation $\cal R$  in the model that glues two linear potentials of different slopes, originally proposed by Starobinsky.
	We find that the enhanced power spectrum reaches its maximum at the wavenumber which is $\pi$ times the junction scale. The peak is $\sim2.61$ times larger than the ultraviolet plateau. We also show that its near-peak behavior can be well approximated by a constant-roll model, once we define the effective ultra-slow-roll $e$-folding number appropriately by considering the contribution from non-single-clock phase only.
Such an abrupt transition to non-attractor phase can leave some interesting characteristic features in the energy spectrum of the scalar-induced gravitational waves, which are detectable in the space-borne interferometers if the primordial black holes generated at such a high peak are all the dark matter. 
\end{abstract}
\maketitle


\section{Introduction}

Inflation is the leading paradigm to solve the problems in the traditional big bang cosmology \cite{Starobinsky:1980te,Guth:1980zm,Linde:1981mu,Albrecht:1982wi}. It can also generate and amplify the quantum fluctuations \cite{Starobinsky:1979ty,Mukhanov:1981xt}, which later reenter the Hubble horizon to seed the anisotropic temperature fluctuations in the cosmic microwave background (CMB) radiation and the density perturbations in the large scale structure (LSS). Although the primordial power spectrum of the scalar-type perturbation has been constrained to be of order $10^{-9}$ at high accuracy on scales larger than 1 Mpc~\cite{Akrami:2018odb}, the lack of resolution on small scales from either CMB or LSS allows us to consider an enhanced scalar power spectrum on small scales, which has fruitful phenomena like primordial black hole (PBH) ~\cite{Zeldovich:1967lct,Hawking:1971ei,Carr:1974nx,Meszaros:1974tb,Carr:1975qj,Khlopov:1985jw} and scalar-induced gravitational wave (GW)  \cite{Matarrese:1992rp,Matarrese:1993zf,Matarrese:1997ay,Noh:2004bc,Carbone:2004iv,Nakamura:2004rm,Ananda:2006af,Osano:2006ew}. PBH forms as the consequence of gravitational collapse in the early universe well before the recombination, which could be a good candidate for the cold dark matter. Recent observations narrow down the mass window for PBHs as all the dark matter to be $10^{16}\sim10^{22}$ g, i.e. the asteroid-mass PBHs~\cite{Carr:2016drx,Sasaki:2018dmp,Carr:2020gox,Carr:2020xqk}. The existence and abundance of such asteroid-mass PBHs can be cross-checked by the GWs induced by the scalar perturbation that collapsed into such BHs~\cite{Saito:2008jc,Saito:2009jt,Bugaev:2009zh,Assadullahi:2009jc,Bugaev:2010bb,Cai:2018dig,Bartolo:2018evs}, which are in the millihertz frequency band. This is an important scientific goal of the space-borne interferometers like LISA~\cite{Barausse:2020rsu,LISA:2022kgy}, Taiji~\cite{Hu:2017mde,Guo:2018npi}, and TianQin~\cite{TianQin:2015yph}.

To generate significant amount of PBHs and induced GWs, usually an enhanced power spectrum of the primordial scalar perturbation on small scales is required. Popular mechanisms for the enhancement include non-slow-roll stage in single field inflation~\cite{Starobinsky:1992ts,Ivanov:1994pa,Yokoyama:1998pt,Garcia-Bellido:2016dkw,Cheng:2016qzb,Garcia-Bellido:2017mdw,Germani:2017bcs,Cheng:2018yyr,Dalianis:2018frf,Tada:2019amh,Xu:2019bdp,Mishra:2019pzq,Bhaumik:2019tvl,Liu:2020oqe,Atal:2019erb,Fu:2020lob,Vennin:2020kng,Ragavendra:2020sop,Gao:2021dfi,Cai:2022erk,Karam:2022nym}, 
 multi-field inflation~\cite{GarciaBellido:1996qt,Kawasaki:1997ju,Frampton:2010sw,Giovannini:2010tk,Clesse:2015wea,Inomata:2017okj,Gong:2017qlj,Inomata:2017vxo,Espinosa:2017sgp,Kawasaki:2019hvt,Palma:2020ejf,Fumagalli:2020adf,Braglia:2020eai,Anguelova:2020nzl,Romano:2020gtn,Gundhi:2020zvb,Gundhi:2020kzm,Cai:2021wzd,Ishikawa:2021xya,Spanos:2021hpk,Hooshangi:2022lao},  
 modified gravity~\cite{Kannike:2017bxn,Pi:2017gih,Gao:2018pvq,Cheong:2019vzl,Cheong:2020rao,Fu:2019ttf,Dalianis:2019vit,Lin:2020goi,Fu:2019vqc,Aldabergenov:2020bpt,Aldabergenov:2020yok,Yi:2020cut,Gao:2020tsa,Dalianis:2020cla,Gao:2021vxb,Wu:2021zta,Teimoori:2021thk,Chen:2021nio,Kawai:2021edk,Zhang:2021rqs,Yi:2022anu,Zhai:2022mpi,Cheong:2022gfc}, 
 curvaton scenario~\cite{Kawasaki:2012wr,Kohri:2012yw,Ando:2017veq,Ando:2018nge,Pi:2021dft}, 
parametric resonance~\cite{Cai:2018tuh,Chen:2020uhe,Chen:2019zza,Cai:2019jah,Cai:2019bmk}, 
oscillons~\cite{Cotner:2016cvr,Cotner:2017tir,Cotner:2018vug,Cotner:2019ykd}, etc. Among all of these, in the single-field inflation, the non-slow-roll stage can be realized by a segment of flat potential \cite{Leach:2000yw,Tsamis:2003px,Kinney:2005vj}. In the ordinary canonical single-field inflation, inflaton $\phi$ rolls down slowly along a potential $V(\phi)$. As the power spectrum from the slow-roll stage is inversely proportional to the first slow-roll parameter $\epsilon\equiv-\dot H/H^2$, it is straightforward to see that a near-flat segment of the potential can generate smaller $\epsilon$ than that of the CMB scales, which can enhance the power spectrum even at slow-roll level. However, such estimation has a fatal flaw, as the ``decaying mode''  (which is decaying on superhorizon scales thus negligible in the slow-roll case) now becomes growing on superhorizon scales, which can dominate the power spectrum around the transition. Consequently, the growing mode can not only greatly change the growth rate of the power spectrum, from a naive estimation of $(H^4/\dot\phi^2)\sim k^6$ to the ``steepest growth'' of $k^4$~\cite{Byrnes:2018txb,Carrilho:2019oqg,Ozsoy:2019lyy,Tasinato:2020vdk,Cole:2022xqc}, but may also leave modulated oscillations due to the mode mixing around the transition scale \cite{Namjoo:2012xs,Pi:2019ihn}, when the transition is sharp. The sharp transition can be well approximated by described by glueing two linear potentials with different slopes at a certain point $\phi_0$. This model was proposed by Starobinsky~\cite{Starobinsky:1992ts}, and was soon applied to generate PBHs~\cite{Ivanov:1994pa}. Ref.\cite{Leach:2001zf} studied the Starobinsky model by the so-called gradient expansion method, while the Wands duality \cite{Wands:1998yp} was identified. It was found that the superhorizon evolution of the growing mode is essential to forge the final shape of the power spectrum. By this method,  the $k^4$-growth of the power spectrum can be explained by the leading-order contribution from the superhorizon evolution of the growing mode, as $\mathcal{P}_\mathcal{R}\sim\mathcal{R}^2\sim\left(\nabla^2\right)^2\sim k^4$~\cite{Ozsoy:2019lyy}. Besides, the mixing of positive- and negative-modes in the new phase superimposes modulated oscillations in addition to the background enhancement, of which the first crest gives the maximum of $\mathcal{P}_\mathcal{R}$ \cite{Biagetti:2018pjj}.

When considering the PBH mass function, the accurate position and height of the peak is of crucial importance, as the PBH abundance is quite sensitive to the amplitude and shape of the power spectrum. This motivates us to revisit the Starobinsky model. We calculate the power spectrum by matching the solutions of Mukhanov-Sasaki equation, and by using the gradient expansion method. After obtaining the analytical formula, we take its large enhancement limit (i.e. the ratio of the slow-roll parameters before and after the transition is large, $\epsilon_+/\epsilon_-\gg1$). As the period of the modulated oscillation is $\pi$, we find that the first crest of the power spectrum locates at $\pi k_0$, where $k_0$ is the wavenumber of which the mode exits the horizon at $\phi_0$. Then the extra enhancement of $\mathcal{P}_\mathcal{R}$ caused by the modulation can be calculated analytically, which is a pure number of $1+15/\pi^2+9/\pi^4\approx2.61$, independent of any parameter as long as $\epsilon_+/\epsilon_-\gg1$. For the PBH mass function $f_\text{PBH}(M)$, if PBHs are all the dark matter, of which the central mass of the peak $M_c$ is in the asteroid-mass window, we find that the peak value $f_\text{PBH}(M_c)\sim\mathcal{O}(1)$ is ``only'' $\sim8$ orders of magnitude larger than that of the UV plateau, $f_\text{PBH}(M\ll M_c)$. In such a case, the small-mass tail of the PBH mass function generates too many tiny PBHs of $10^{14}~\text{g}$, an inevitable byproduct of the Starobinsky model when $\int f_\text{PBH}(M)\mathrm{d}\ln M=1$. This contradicts with the null detection of extra-galactic gamma ray from their Hawking radiation. A possible solution to this problem might be an extra suppression of the UV plateau, for instance $\grave{\textit{a}}$ \textit{la} Ref.\cite{Ivanov:1994pa}, a prototype of the near-inflection point inflation well studied recently. 

We also compare the Starobinsky model to the constant-roll inflation, parameterized by simply changing the second slow-roll parameter $\eta\equiv\dot\epsilon/(H\epsilon)$ instantaneously from 0 to $-6$, and then back to $\sim0$ after a few $e$-folds. We find that the $e$-folding number for the USR stage in the constant-roll model is crucial. We propose a new definition of effective $e$-folding number, which can generate the same enhancement including the modulations in the constant-roll inflation. 

This paper is organized as follows. In section \ref{s:model} we review the Starobinsky model, and study the power spectrum both numerically and analytically, by matching the solution of Mukhanov-Sasaki equation and by using the gradient expansion method. We study the characteristic features in the shape of the power spectrum obtained, and compare it with the one we get in the constant-roll inflation in Section \ref{s:shape}. In Section \ref{s:pbh} we discuss PBHs and induced GWs. We conclude in Section \ref{s:conclusion}, with a short discussion on the effect of non-Gaussianity and quantum diffusion. Appendix \ref{a:model} is a short review of the background dynamics of Starobinsky model. Appendix \ref{a:coefficient} displays the formulas of how to match the solution at $\phi_0$. In Appendix \ref{a:alpha}, we review the gradient expansion method, and show how to calculate the superhorizon enhancement factor. We list some functions in constant-roll model in Appendix \ref{a:constantroll}.

\section{Power spectrum in the Starobinsky Model}\label{s:model}
\begin{figure}
\begin{center}
\includegraphics[width=0.49\textwidth]{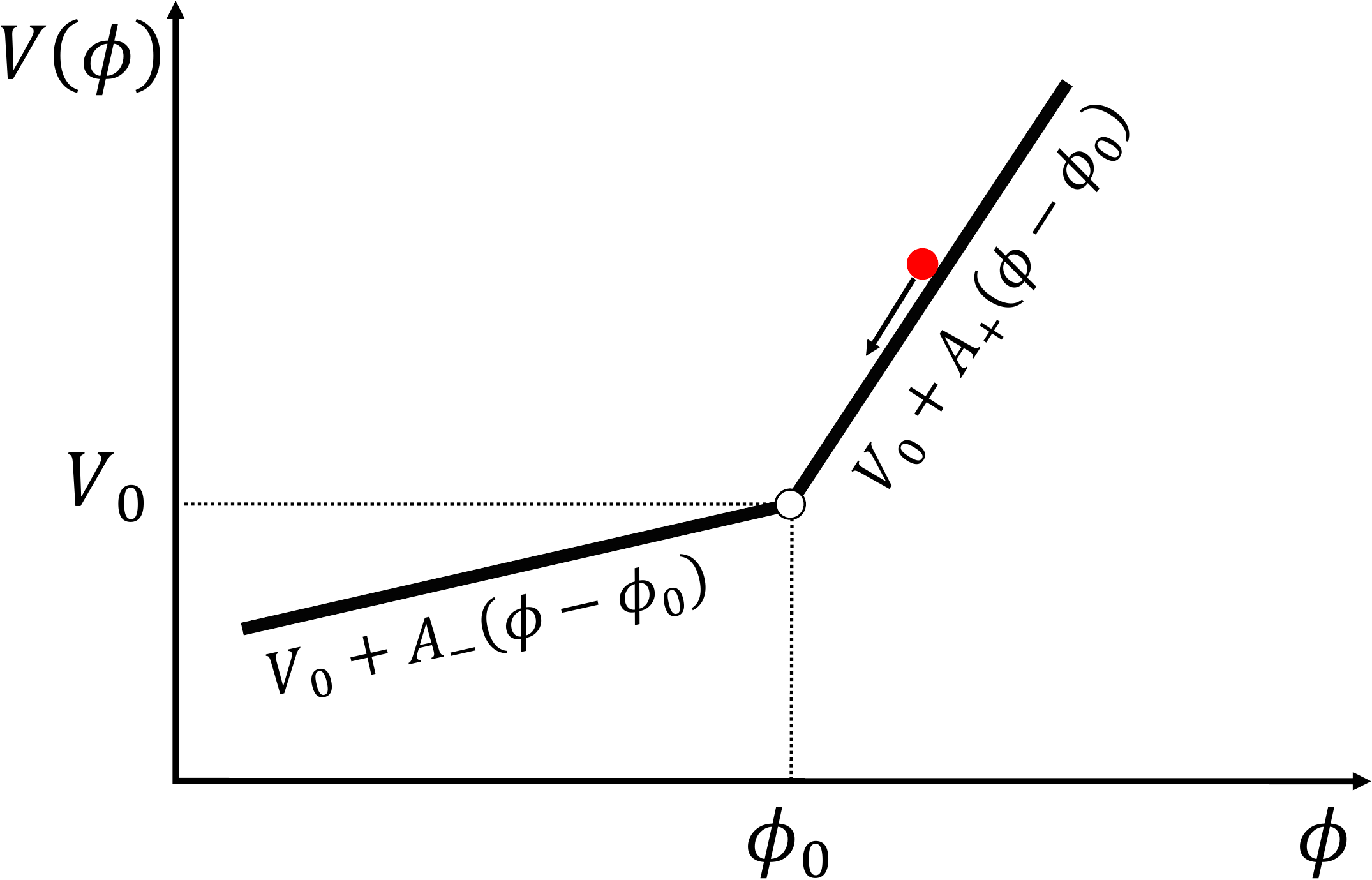}
\includegraphics[width=0.49\textwidth]{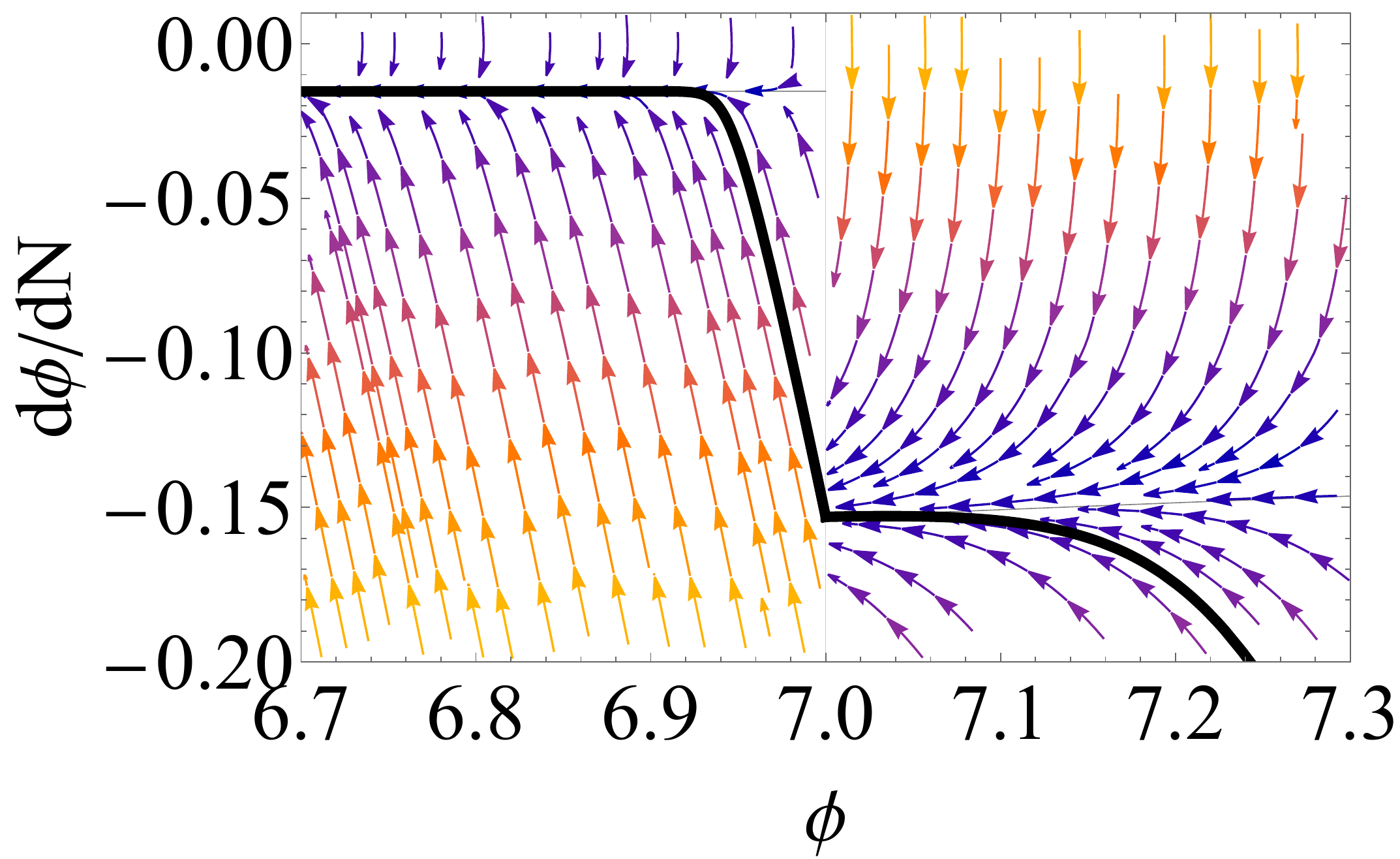}
\caption{Left: A schematic picture of $V(\phi)$ in Starobinsky model. Right: The phase portrait of Starobinsky model, with $\phi_0=7$, $\Lambda \equiv A_+ / A_-=10$, $V_0=3 H_0^2$ ($\mpl$ is set to be 1).}
\label{f:schematic}
\end{center}
\end{figure}

In the Starobinsky model, the canonical inflaton field rolls down a piecewise linear potential~\cite{Starobinsky:1992ts}:
\be\label{themodel}
V(\phi)= \begin{cases}V_{0}+A_{+}\left(\phi-\phi_{0}\right), & \text { for } \phi>\phi_{0}; \\ V_{0}+A_{-}\left(\phi-\phi_{0}\right), & \text { for } \phi\leq \phi_{0}.\end{cases}
\ee
$\phi_0$ is called a ``singularity'' in Ref.\cite{Starobinsky:1992ts}, in the sense that $V''(\phi_0)$ is vary large. The potential \eqref{themodel} is a simplified version of the original potential proposed in Ref.\cite{Starobinsky:1992ts} where a smoothing scale $\Delta\phi\ll\mpl$ is introduced such that the field velocity is continuous. \footnote{Ref.\cite{Starobinsky:1992ts} also considered another model with a velocity loss/boost at the singularity. This will bring strong non-Gaussianity which is recently studied in Ref.\cite{Cai:2022erk}.} This smoothing scale, however, can be ignored if we keep in mind that the field velocity well before and after the singularity is continuous, as $\Delta\phi\ll\mpl\ll\phi_0$. A globally linear potential for canonical single-field inflation has been excluded by Planck/BICEP2/Keck on CMB scales \cite{BICEP:2021xfz}, so throughout this paper we only study the local dynamics around this singularity, which means $V_0\gg A_{\pm}(\phi-\phi_0)$ and the Friedmann equation is well approximated by
\be\label{def:H0}
H^2=\frac{V(\phi)}{3\mpl^2}\approx\frac{V_0}{3\mpl^2}\equiv H_0^2.
\ee
A schematic picture of the potential $V(\phi)$ is shown in the left panel of Fig. \ref{f:schematic}. 

In Ref.\cite{Starobinsky:1992ts}, the scale of $\phi_0$ was put right below the CMB scale ($\sim1~\text{Mpc}$) with $A_-/A_+\sim3$, as its physical motivation is mainly to explain a modest suppression of the density perturbation toward smaller scales. Recently, this model was used the other way round ($A_-/A_+\sim0.1$ at $k_0^{-1}\sim10^4$~Mpc) to explain the low-$\ell$ anomaly, i.e. the deficiency of power at large scales \cite{Akrami:2018odb,Sinha:2005mn,Hazra_2021}, we put $k_0\sim10^{12}~\text{Mpc}^{-1}$ and $A_-/A_+\ll1$, such that the enhanced power spectrum can generate PBHs as all dark matter. 

Before that, let us study the general properties of the power spectrum in the Starobinsky model. For simplicity, let us define the ratio of the slopes as
\be\label{def:Lambda}
\Lambda\equiv\frac{A_+}{A_-}.
\ee
The initial condition of the inflaton $\phi$ in the first stage ($\phi>\phi_0$) should be on the slow-roll attractor, with the slow-roll parameter
\be
\epsilon_+\equiv\left.-\frac{\dot H}{H^2}\right|_+=\frac{A_+^2}{18H_0^4\mpl^2},
\ee
which gives the power spectrum of the first stage
\be\label{PIR}
\mathcal{P}_{\mathcal{R}}^{\mathrm{(IR)}}=\frac{H_0^2}{8\pi^2\epsilon_+\mpl^2}=\frac{9}{4 \pi^{2}} \frac{H_{0}^{6}}{A_{+}^{2}}.
\ee
In reality, the value of the IR plateau of the power spectrum should be consistent with that of the CMB scales. An extrapolation of the primordial power spectrum to small scales could be quite different from that on CMB scales, when a large running ($\dif n_s/\dif\ln k$) and running of running ($\dif^2n_s/\dif\ln k^2$) are considered \cite{Green:2018akb}. In this paper, just for simplicity, we only consider a constant spectral tilt, which extrapolates the power spectrum linearly to a small scale $k_\text{IR}=10^{10}~\text{Mpc}^{-1}$ as
\be
\mathcal{P}_\mathcal{R}^\text{(IR)}= A_*\left(\frac{k_\text{IR}}{k_*}\right)^{n_s-1}\approx8.5\times10^{-10},
\ee
where in the second step we set $A_*=2.11\times10^{-9}$ and $n_s=0.965$ at the pivot scale $k_*=0.05~\text{Mpc}^{-1}$ \cite{Akrami:2018odb}. Here $k_\text{IR}$ is two orders smaller than the wavenumber of asteroid-mass PBH, $k_0=10^{12}~\text{Mpc}^{-1}$, which can be set as the scale for an initial infrared power spectrum well before the enhancement.

\subsection{Matching the Mukhanov-Sasaki equation}

A smooth transition to the other linear potential with slope $A_-$ of $\phi<\phi_0$ set the velocity of the previous slow-roll attractor as the initial condition at $\phi_0$, which is not the slow-roll velocity of the second stage, thus a transient non-attractor phase appears. By the virtue of $e$-folding number
\be
N\equiv\int^t_{t_0} H\dif t\approx H_0(t-t_0),
\ee
a schematic phase portrait of $\phi(N)$ is shown in the right panel of Fig.\ref{f:schematic}. The equation of motion for the curvature perturbation on comoving hypersurfaces, $\mathcal{R}$, can be written as
\be
\mathcal{R}_{k}^{\prime \prime}+2 \frac{z^{\prime}}{z} \mathcal{R}_{k}^{\prime}+k^{2} \mathcal{R}_{k}=0,
\label{R}
\ee
where $z\equiv a M_{\mathrm{Pl}} \sqrt{2\epsilon}$. A prime denotes the derivative with respect to the conformal time
\be
\tau=-\int^{t_f}_{t}\frac{\dif t'}{a(t')}=-\frac1{a_0H_0}e^{-H_0(t-t_0)}=\tau_0e^{-H_0(t-t_0)},
\ee
where $t_f\gg H_0^{-1}$ is the end of inflation. We normalize $\tau$ at $\tau_0=-1/(a_0H_0)$, where $a_0$ and $H_0$ are the scale factor and Hubble parameter at $\phi=\phi_0$. By solving the background equation of motion for $\phi$ with the slow-roll initial condition, we can write down the analytical form of $z(\tau)=a\dot\phi/H$ near $\phi_0$:
\be
z(\tau)= \begin{cases}\displaystyle-\dfrac{a_{0}}{3 H_{0}^{2}} A_{+}\left(\frac{\tau}{\tau_{0}} \right)^{-1}, & \text { for } \tau<\tau_{0}; \\ 
\\
\displaystyle-\dfrac{a_{0}}{3 H_{0}^{2}}\left[A_{-}\left(\frac{\tau}{\tau_{0}} \right)^{-1}+\left(A_{+}-A_{-}\right)\left(\frac{\tau}{\tau_{0}} \right)^{2}\right], & \text { for }\tau\geq \tau_{0}.
\end{cases}
\label{zz}
\ee
The details of deriving \eqref{zz} are written in Appendix \ref{a:model}. In the transient non-attractor phase, the $\tau^2$-term in \eqref{zz} is dominant, which is equivalent to a collapsing dust-dominated universe ($z\propto a\propto\tau^2$), dubbed Wands duality~\cite{Wands:1998yp}.
To solve it analytically, we define $v\equiv z\mathcal{R}$ and write down the Mukhanov-Sasaki equation
\be\label{MSeqn}
v_k''(\tau)+\left(k^2-\frac{z''}{z}\right)v_k(\tau)=0.
\ee
The effective potentials of the mode function $v_k(\tau)$ degenerate in both phases, as $z''/z=2/\tau^2$, so the solutions can be expressed by Hankel function of order 3/2 with different coefficients:
\be
v_{k,i}(\tau)=C_i(k)   \sqrt{-k \tau}   H_{3 / 2}^{(1)}(-k \tau)+D_i(k)   \sqrt{-k \tau}   H_{3 / 2}^{(2)}(-k \tau) , 
\label{solvk}
\ee
where $i=1,~2$ for the first ($\tau<\tau_0$) and second ($\tau>\tau_0$) stage, respectively. 
For the first stage, the coefficients $C_1$ and $D_1$ are set by the Bunch-Davies vacuum state at an initial time when all the related lengths are deep inside the horizon~\cite{Bunch:1978yq}. Thanks to the sudden transition, the solution in the second stage can be easily achieved by appropriately matching the solutions at $\tau_0$: the coefficients $C_2$ and $D_2$ are determined by the Israel boundary condition~\cite{Israel:1966rt,Deruelle:1995kd} that both $\mathcal{R}$ and $\mathcal{R}'$ are continuous at $\tau_0$. The details are shown in Appendix \ref{a:coefficient}. We will see that as the vacuum changes instantaneously, the negative-frequency modes in the new vacuum will be excited transiently, which causes the modes to mix and superimposes modulated oscillations on the enhanced/suppressed step-like power spectrum. 

After getting $C_2$ and $D_2$, we have an expression for the solution when $\tau>\tau_0$. Taking its $\tau\to 0^-$ limit by using the asymptotic form of the Hankel function
\be
H_{3 / 2}^{(1)}(x \rightarrow 0) \longrightarrow-i   \sqrt{\frac{2}{\pi}}   \frac{1}{x^{3 / 2}},
\ee
we can obtain the analytical formula for the power spectrum in the Starobinsky model at the end of inflation:
\be
\frac{\mathcal{P}_{\mathcal{R}}\left(\kappa\right)}{ \mathcal{P}_{\mathcal{R}}^{\mathrm{(IR)}}}= \frac{9\left(\Lambda-1\right)^2}{\kappa^6}\big(\sin\kappa-\kappa \cos\kappa\big)^4+\left\{ \Lambda +\frac{3\left(\Lambda-1\right)}{2\kappa^3}\left[ (\kappa^2-1)\sin2\kappa+2\kappa\cos2\kappa\right]\right\}^2,
\label{main}
\ee
where $\kappa\equiv-k\tau_0$ is the dimensionless wavenumber normalized by $k_0\equiv-1/\tau_0$, and $\Lambda\equiv A_+/A_-$ is defined in \eqref{def:Lambda}. This result is consistent with what \cite{Starobinsky:1992ts} obtained. 

The infrared (IR) power spectrum is given in \eqref{PIR}, which together with $\Lambda$ controls the shape and amplitude of the power spectrum. The ultraviolet (UV) power spectrum can be obtained by taking the $\kappa\to\infty$ limit of \eqref{main},
\be
\mathcal{P}_\mathcal{R}^\text{(UV)}=\lim_{\kappa\to\infty}\mathcal{P}_\mathcal{R}(\kappa)=\Lambda^2\mathcal{P}_\mathcal{R}^\text{(IR)}=\frac{9}{4 \pi^{2}} \frac{H_{0}^{6}}{A_{-}^{2}}.
\ee
Note that we do not impose any approximation in deriving \eqref{main}, which is thus valid for any value of $\Lambda$, including $\Lambda<1$, i.e. the power spectrum is suppressed on smaller scales as the original Starobinsky model \cite{Starobinsky:1992ts}. For the physical case we considered, the power spectrum should be greatly enhanced to generate enough PBHs, which implies $\Lambda\gg1$. In this case we can expand \eqref{main} for large $\Lambda$. Up to the leading order, a simpler form
\be
\frac{\mathcal{P}_{\mathcal{R}}\left(\kappa\right)}{ \mathcal{P}_{\mathcal{R}}^{\mathrm{(UV)}}}
=1+\frac{9}{2\kappa^2}+\frac{9}{\kappa^4}+\frac{9}{2\kappa^6}+\left(\frac{21}{2\kappa^2}-\frac{9}{2\kappa^6}\right)\cos2\kappa+\left(\frac3\kappa-\frac{12}{\kappa^3}-\frac9{\kappa^5}\right)\sin2\kappa.
\label{Pfapp}
\ee
can well approximate the behavior around the enhanced peaks. An overall $\Lambda^2$ factor has been absorbed in $\mathcal{P}_{\mathcal{R}}^{\mathrm{(IR)}}$, which now becomes $\mathcal{P}_{\mathcal{R}}^{\mathrm{(UV)}}$. This result is consistent with Ref.\cite{Biagetti:2018pjj}, 

\begin{figure}
\begin{center}
\includegraphics[width=0.75\textwidth]{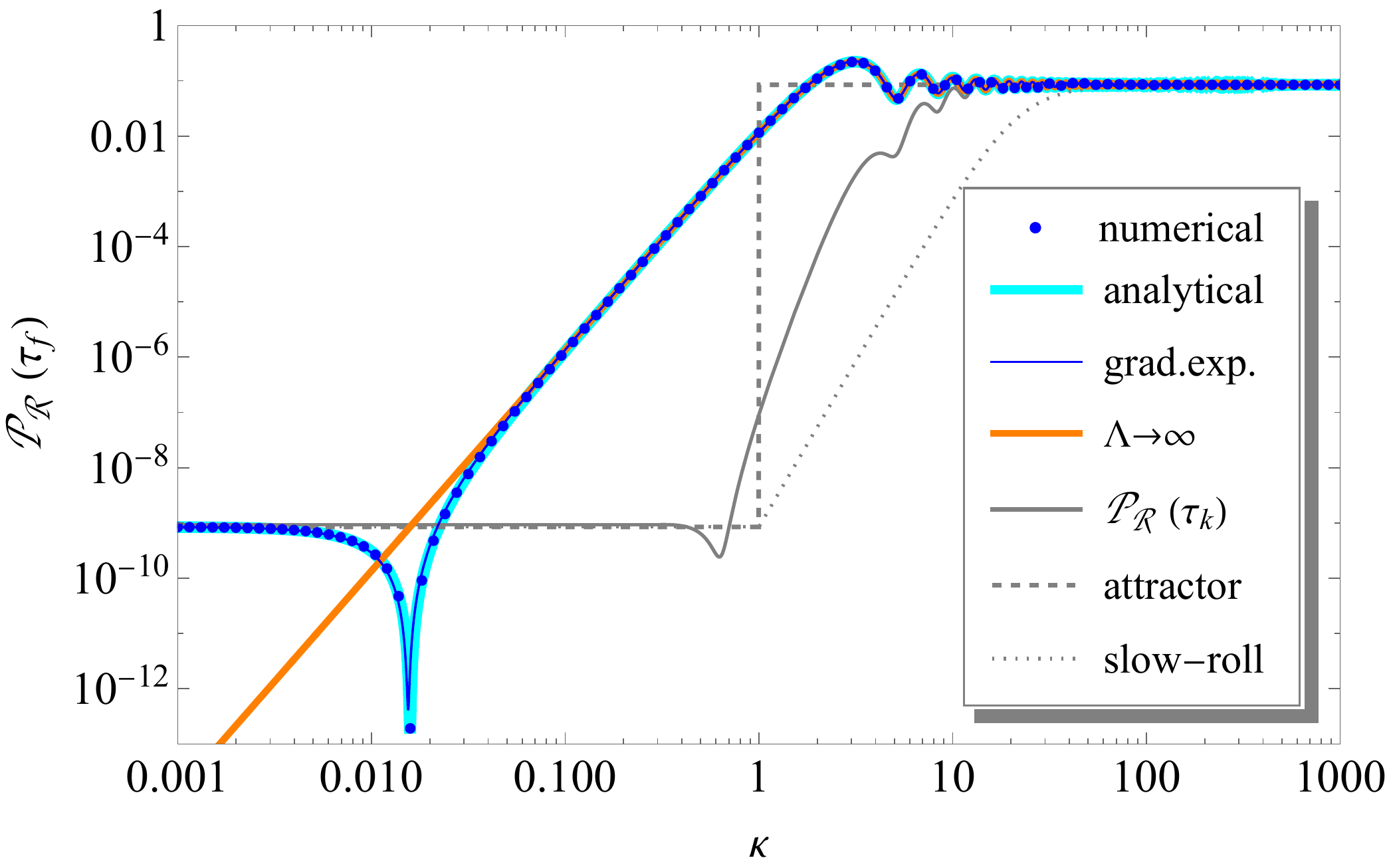}
\caption{The enhanced power spectrum at the end of inflation with $\mathcal{P}_\mathcal{R}^\text{(IR)}=8.5\times10^{-10}$ and $\Lambda=10^4$, calculated by numerically solving the equation of motion \eqref{R} (blue dots), by analytical results of matching the solutions  (cyan thick, see \eqref{main}), by gradient expansion method with $\varrho=\sqrt{0.1}$ (blue curve, see \eqref{gradient}), and by the $\Lambda\to\infty$ approximation (orange thick, see \eqref{Pfapp}). The slow-roll (gray dotted, see \eqref{PRSR}) and step-like attractor (gray dashed) estimations are also shown for comparison, though they do not represent any realistic power spectrum. By comparing with the power spectrum at the horizon exit (gray curve, see \eqref{PRk}), we can easily see that the power spectrum near the maximum are mainly contributed by the superhorizon evolution.}
\label{power spectrum}
\end{center}
\end{figure}

\subsection{The Gradient-Expansion Method}

There is another method based on the gradient expansion to study the Starobinsky model~\cite{Leach:2001zf}, which is quite helpful in understanding the physical origin of the enhancement. The essence of this method is to factorize the power spectrum at the end of inflation  $\mathcal{P}_\mathcal{R}\left(\tau_{f}\right)$ into a superhorizon enhancement factor $\left|\alpha_k\right|^2$ and the power spectrum at horizon exit $\mathcal{P}_\mathcal{R}(\tau_k)$
\be
\mathcal{P}_\mathcal{R}\left(\tau_{f}\right)=\left|\alpha_{k}\right|^{2}   \mathcal{P}_\mathcal{R}\left(\tau_{k}\right).
\label{gradient}
\ee
For the mode function of a given wavenumber $k$, the time of ``horizon exit'' can only be determined up to a factor of order $1$. Here we reflect this uncertainty by setting the horizon exit to be at $\tau_k=-\varrho/k$, which is equivalent to $k \tau_{k}=-\varrho$ and $k / \mathcal{H}_{k}=\varrho$. The power spectrum at the horizon exit $\tau_k$ is
\be
\mathcal{P}_{\mathcal{R}}\left(\tau_{k}\right)=
\calP_\calR^\text{(IR)}\cdot
\begin{cases}1+\varrho^{2} & \text { for } \kappa<\varrho, \\ 
\\
\frac{\displaystyle\mathcal{O}_{0}+\mathcal{O}_{1}\Lambda^{-1}+\mathcal{O}_{2}\Lambda^{-2}}
{\displaystyle2\left[\left(\Lambda^{-1}-1\right) \varrho^{3}-\kappa^{3}\Lambda^{-1}\right]^2}& \text { for }\kappa \geq \varrho, \end{cases}
\label{PRk}
\ee
while the coefficients $\mathcal{O}_{0,1,2}$ are functions of $\varrho$ and $\kappa$, written explicitly in Appendix \ref{a:coefficient}. Here $\kappa<\varrho$ ($\kappa>\varrho$) means the $k$-mode exit the horizon before (after) $\tau_0$. 

The superhorizon enhancement factor is given by
\be\label{def:alpha}
\alpha_{k} \approx 1+\frac{\mathcal{R}_{k}^{\prime}\left(\tau_{k}\right)}{3 \mathcal{H}_{k} \mathcal{R}_{k}\left(\tau_{k}\right)} D_{k}-F_{k},
\ee
where $D_k$, $F_k$ are determined by the evolution of $z(\tau)$ out of the horizon
\begin{align}\label{Ddir}
D_k&\approx3 \mathcal{H}_{k}z^{2}\left(\tau_{k}\right) \int_{\tau_k}^{0^{-}} \frac{\mathrm{d} \tau^{\prime}}{z^{2}\left(\tau^{\prime}\right)}=
\begin{cases}\displaystyle\left(\Lambda-1\right)\left(\frac{\kappa}{\varrho} \right)^{3}+1, & \text { for } \kappa<\varrho; \\ 
\\
\displaystyle\left(\Lambda-1\right)\left(\frac\varrho\kappa\right)^{3}+1, & \text { for }\kappa \geq \varrho. \end{cases} \\\nn
F_k&\approx k^{2} \int_{\tau_k}^{0^{-}} \frac{\mathrm{d} \tau^{\prime}}{z^{2}\left(\tau^{\prime}\right)} \int_{\tau_{k}}^{\tau^{\prime}} z^{2}\left(\tau^{\prime \prime}\right) \mathrm{d} \tau^{\prime \prime}\\
&= \begin{cases}\displaystyle\varrho^{2}\left\{\dfrac{1}{6}+\dfrac{2}{5}\left(\Lambda-1\right)\left(\frac\kappa  \varrho\right)^{2}-\dfrac{1}{3}\left(\Lambda-1\right)\left(\frac\kappa\varrho \right)^{3}\right\}, & \text { for } \kappa<\varrho; \\ 
\\\displaystyle
\varrho^{2}\left\{ \dfrac{1}{6}+\dfrac{1}{15}\left(\Lambda-1\right)\left(\frac\varrho\kappa \right)^{3}\right\}, & \text { for }\kappa\geq \varrho.\end{cases}
\label{Fdir}
\end{align}
Here in the second step we use the expressions for $z(\tau)$ in Starobinsky model given in \eqref{zz}. We can see from \eqref{def:alpha} and \eqref{Fdir} that at the leading order on superhorizon scales, $\mathcal{P_R}\propto|\alpha_k|^2\sim F_k^2\propto k^4$, which gives a straightforward explanation for the universal ``steepest growth'' found in Ref.\cite{Byrnes:2018txb}. Then we get the enhancement factor
\be
\left|\alpha_{k}\right|^{2}=\begin{cases} s_{0}+s_{1} \cdot \Lambda+s_{2} \cdot \Lambda^{2}, & \text { for } \kappa<\varrho; \\ 
\\
\dfrac{1}{900 \kappa^{6}}\left\{W_{1}-\dfrac{W_{2} \cdot \mathcal{W}}{W_{0}}+25 \cdot \dfrac{W_{3}+\mathcal{W}^{2}}{W_{0}^{2}}\right\}, & \text { for }\kappa \geq \varrho.\end{cases}
\label{alpha}
\ee
$s_{0,1,2}$, $W_{0,1,2,3}$ and $\mathcal{W}$ are functions of $\kappa$, $\varrho$, and $\Lambda$, which are written explicitly in Appendix \ref{a:alpha}. 
By multiplying \eqref{PRk} and \eqref{alpha}, we get the power spectrum given by the gradient expansion method, which slightly depends on $\varrho$ when $\varrho\lesssim\mathcal{O}(1)$. In Fig.\ref{power spectrum} we show our result, following the convention of Ref.\cite{Leach:2001zf} that $\varrho=\sqrt{0.1}$. It can be compared with the analytical result \eqref{main}, which is simply the $\varrho \to 0$ limit of \eqref{gradient}, because the result of directly solving the equation of motion should be recovered by pushing the moment of ``horizon exit'' $\tau_k$ to the end of inflation, while the superhorizon enhancement factor $|\alpha_k|^2$ degenerates to $1$. 


\begin{figure}
\begin{center}
\includegraphics[width=0.75\textwidth]{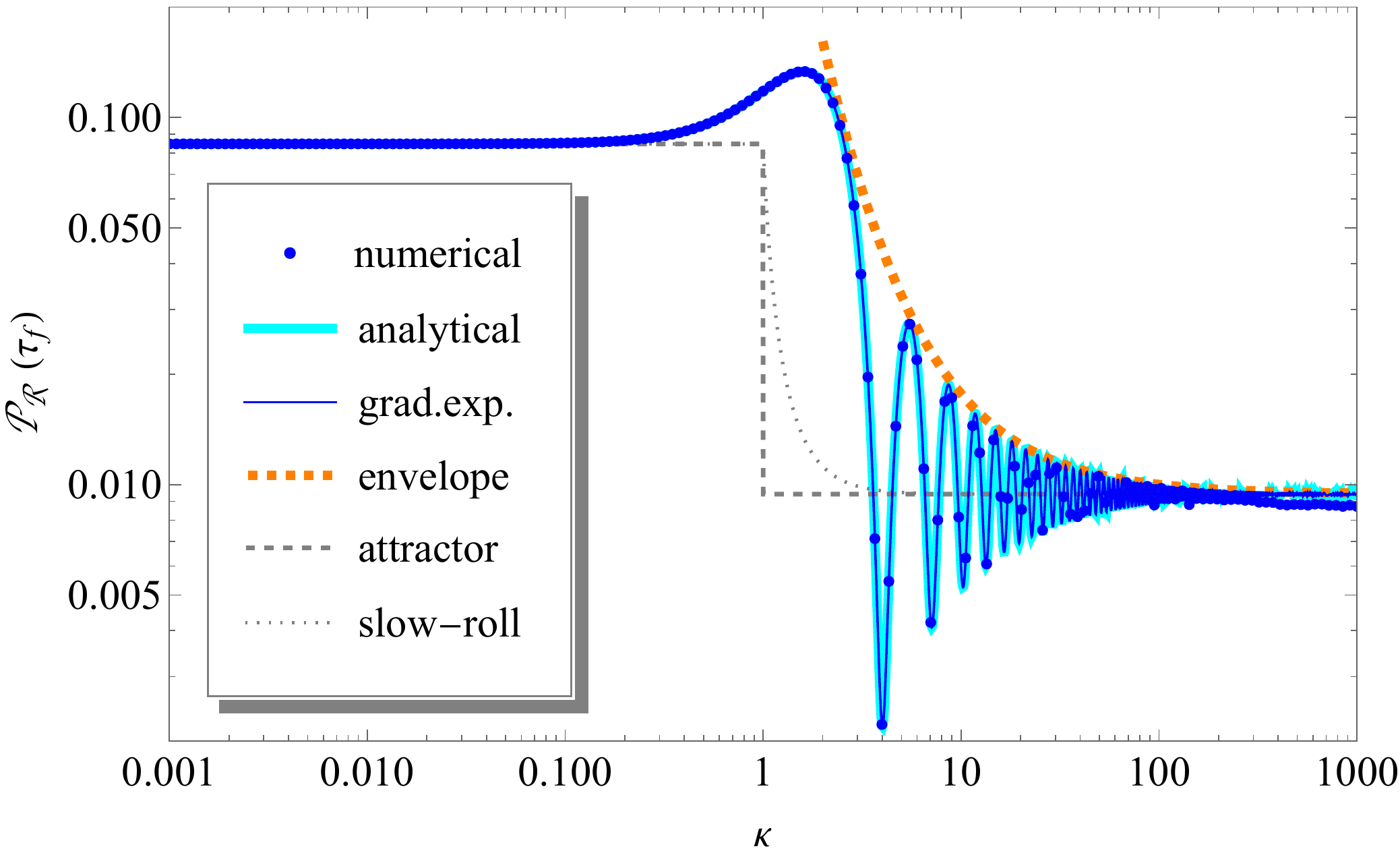}
\caption{Step-down Starobinsky model with $\mathcal{P}_\mathcal{R}^\text{(IR)}=8.5\times10^{-2}$, $\Lambda=1/3$. Blue dots: numerical solution of \eqref{R}; cyan curve: analytical solution \eqref{main}; blue curve: gradient expansion result (\ref{gradient}) with $\varrho=\sqrt{0.1}$; gray dotted: slow-roll solution (\ref{PRSR}); gray dashed: step-like attractor; thick orange dashed: the envelope of oscillations when exiting from USR \eqref{envelope}.}
\label{power spectrum nonPBH}
\end{center}
\end{figure}

\subsection{Comparison}

All of these results are shown in Fig.\ref{power spectrum} for comparison, including a naive estimation by the slow-roll solutions in both stages. To do so, note that the comoving curvature perturbation $\mathcal{R}$ is connected to the field perturbation on the spatially-flat slicing, $Q\equiv\delta\phi^\text{(flat)}$, by
\be
\mathcal{R}=-\frac{H}{\dot{\phi}}Q.
\ee
The quantum fluctuation of a massless scalar field has a variance of $H^2/(4\pi^2)$ at the horizon crossing, which means the power spectrum of the comoving curvature perturbation in the slow-roll formula is
\be\label{PRSR}
\mathcal{P}_\mathcal{R}^\text{(SR)}=\left.\frac{H_0^2}{\dot\phi^2}\langle Q^2\rangle\right|_{\tau_k}=\frac{H_0^4}{4\pi^{2}\dot{\phi}\left(\tau_k\right)^2}.
\ee
Considering that the field velocity $\dot{\phi}$ in the Starobinsky model is
\be
\dot{\phi}(\tau_k)= \begin{cases}-\dfrac{A_+}{3H_0}, & \text { for } \kappa<1, \\ 
\\
-\dfrac{A_- + \left( A_+ -A_- \right)\kappa^{-3}}{3H_0}, & \text { for } \kappa \geq 1,\end{cases}
\ee
we can immediately see that power spectrum obtained by slow-roll solutions which incorrectly considers the constant mode only can roughly give the transition between the two attractors. But there are a few flaws: (1) the growth rate is $H^4/\dot\phi^2\sim k^6$, which is faster than the steepest growth rate $k^4$; (2) the growth happens at a larger wavenumber, which reflects its ignorance of the superhorizon growth; and (3) the modulated oscillation originates from the transient excitation of negative-frequency modes is absent. 

For completeness, we also consider the original parameter choice by Starobinsky in \cite{Starobinsky:1992ts} with $A_+<A_-$, which we will call step-down Starobinsky model for short. Obviously, this potential can suppress the power spectrum but also superimpose modulated oscillations. The envelope of such a damped oscillation can be easily obtained from \eqref{main}:
\begin{align}
\frac{\mathcal{P}_{\mathcal{R}}^{\mathrm{(env)}}}{\mathcal{P}_{\mathcal{R}}^{\mathrm{(IR)}}}=&
\frac{9 (\Lambda -1)^2}{\kappa ^5}+\frac{9 (\Lambda -1)^2}{\kappa ^4}
+\frac{3 \left(4 \Lambda ^2-7 \Lambda +3\right)}{\kappa ^3} 
+\frac{3 \left(5 \Lambda ^2-8 \Lambda +3\right)}{\kappa ^2}-\frac{3 \Lambda  (\Lambda -1)}{\kappa }+\Lambda ^2 .
\label{envelope}
\end{align}
The power spectrum as well as the envelope in the step-down Starobinsky model is shown in Fig.\ref{power spectrum nonPBH}.


\section{A detailed study on the Characteristic features}\label{s:shape}

In this section we study the shape of the power spectrum enhanced by the Starobinsky model, especially (1) the position of the peak(s), (2) the maximal power spectrum, (3) the dip before enhancement, and (4) its equivalence to an appropriately parameterized constant-roll model. 



\subsection{Peak from the oscillation}
From \eqref{main} and \eqref{Pfapp}, we can see that because of the transient excitation of the negative-frequency modes, the power spectrum has modulated sinusoidal oscillations with period $\pi$. The approximate analytical expression (\ref{Pfapp}) can describe the behavior on the plateau of the power spectrum quite well. Taking the extreme values of the power spectrum, i.e. $\calP_\calR'(\kappa)=0$, we have
\begin{align}
&\sin \left(2 \kappa_c-\Psi\right)=-\frac{3 \left(\kappa ^4+4 \kappa ^2+3\right)}{ \sqrt{ \left(\kappa ^2+1\right) \left(4 \kappa ^{10}+9 \kappa ^6+63 \kappa ^4+135 \kappa ^2+81\right)}}, \\
&\Psi=\arctan \left(\frac{2 \kappa ^6-15 \kappa ^4-6 \kappa ^2+9}{8\kappa ^5-12 \kappa ^3-18 \kappa }\right).
\label{pi}
\end{align}
As the amplitude of the oscillation is decreasing, the global maximum of the power spectrum appears at the crest of the first period:
\begin{align}
&2 \kappa_c-\Psi=2 \pi-\Theta, \\ 
&\Theta=\arcsin\left(\frac{3 \left(\kappa ^4+4 \kappa ^2+3\right)}{ \sqrt{ \left(\kappa ^2+1\right) \left(4 \kappa ^{10}+9 \kappa ^6+63 \kappa ^4+135 \kappa ^2+81\right)}} \right).
\label{eq:peakcond}
\end{align}
Interestingly, $\Theta-\Psi\ll1$, which leaves $\kappa_c=\pi$ at leading order. More accurate result can be obtained by iteration, for instance 
\begin{align}
\kappa_c^{(1)} &=\pi+\left.\frac{1}{2}\left(\Theta-\Psi\right)\right|_{\kappa=\pi}\approx\pi-0.002.
\end{align}
Compared with the numerical result $\kappa_c^\text{(num)}=3.1385\cdots$, our result $\kappa_c\approx\pi$ is a good approximation up to $\mathcal{O}(0.1\%)$, which gives
\be
\mathcal{P}_{\mathcal{R}}^{\mathrm{(max)}}\approx\mathcal{P}_{\mathcal{R}}(\kappa=\pi)=\left(1+\frac{15}{\pi^{2}}+\frac{9}{\pi^{4}}\right)\mathcal{P}_\mathcal{R}^\text{(UV)}\approx2.61\mathcal{P}_\mathcal{R}^\text{(UV)}.
\label{nextleading}
\ee
This extra enhancement caused by the first crest of the modulated oscillation is independent of the height of the plateau ($\sim\Lambda^2$), therefore independent of the slow-roll parameter. It is a pure number determined only by the position of the first crest of the oscillations, about $\pi$ times the junction wavenumber $k_0$. Although this peak position is derived in Starobinsky model with a piecewise linear potential, we believe that it also holds when the second slow-roll parameter $\eta$ changes fast enough. Of course, if $\eta$ changes only slowly, there will be no oscillation and the above arguments will not hold \cite{Cole:2022xqc}. In this case, the maximum of the power spectrum appears near the end of the plateau, $\sim\frac\pi2e^{N_\text{USR}}$ (see \eqref{dropscale} below).
\subsection{Dip}

An obvious dip appears in the IR power spectrum before it starts to grow. The position of the dip can be easily obtained by checking the minimum of \eqref{main} when $\kappa\ll1$, which gives
\be\label{dip0}
\kappa_\text{dip}\approx\left(\frac25(\Lambda-1)+\frac67\right)^{-1/2},
\ee
up to $\mathcal{O}(\kappa^4)$. When $\Lambda\gg1$, this result coincides with $\kappa_\text{dip}\approx\sqrt{5/(2\Lambda)}$ firstly found by \cite{Starobinsky:1992ts}.

The appearance of such a dip can be explained as follows. As we know, the solution of $\mathcal{R}$ on superhorizon scales is usually written as a sum of the constant mode and a time-dependent mode:
\be\label{Rsuperhorizon}
\mathcal{R}_{k \to 0}=A_{k}+B_{k} \int^{t} \frac{dt'}{a^{3}(t') \epsilon(t')},
\ee
where we transferred to cosmic time for physical clearance. In the slow-roll case when $\epsilon$ is nearly a constant, the second mode is always decaying. However, when $\epsilon$ decays fast enough, i.e. $\epsilon \propto a^{p}$ with $p<-3$, the second term turns to grow. In the language of the second slow-roll parameter $\eta\equiv\dot\epsilon/(H\epsilon)$, the second mode becomes growing on superhorizon scales when $\eta<-3$, which is called non-single-clock inflation in Ref.\cite{Byrnes:2018txb}. The well known USR inflation, $\eta=-6$, is a special case.

For the modes that leave the horizon deep in the first slow-roll phase, $\mathcal{P}_\mathcal{R}(\tau_k)$ for $\kappa\ll \varrho$ is dominated by the constant mode, which yields a near-flat spectrum (\ref{PRk}). As $\kappa$ increases, the growing mode are more involved in the superhorizon enhancement, which finally becomes comparable with the constant mode and even dominates it. As the result of the competition, a dip appears in the enhancement factor $\left| \alpha_k\right|^2$ and thus in the power spectrum. The position of dip can also be obtained by calculating the minimum of the enhancement factor given in (\ref{alpha}) for $\kappa<\varrho$, and taking the limit that $\varrho \to 0$, 
\be\label{dipposition}
\kappa_{\mathrm{dip}}\approx\sqrt{\frac{5}{2\Lambda}},
\ee
which is exactly what we found in \eqref{dip0}.

\subsection{Comparison with constant-roll}\label{s:constantroll}

It is interesting to compare the power spectrum in the Starobinsky model and the constant-roll model, which displays some similar features, as long as we define the effective $e$-folding number for the USR phase in the latter model appropriately. In constant-roll model, $\eta(N)$ is described by step functions, which goes from 0 to $-6$, lasts for $N_\text{USR}$ $e$-folds, and then goes back to 0:
\be\label{stepfunc}
\eta=-6\Theta\left( \tau_1-\tau\right)\Theta\left( \tau-\tau_0\right),
\ee
where $\Theta$ is the Heaviside step function. 
Integrating \eqref{stepfunc}, we get the first slow-roll parameter
\be
\epsilon= \epsilon_0 \cdot \begin{cases}1, & \text { for } \tau<\tau_{0}, \\ 
\left(\tau / \tau_{0}\right)^{6}, & \text { for } \tau_{0}<\tau<\tau_{1}, \\
 e^{-6N_{\mathrm{USR}}}, & \text { for }\tau>\tau_{1}. \end{cases}
\label{epsilon3}
\ee
We choose $\epsilon_0=\epsilon_+$, such that the two models share the same $\mathcal{P}^{\mathrm{(IR)}}_{\mathcal{R}}$.
If we want to approximate the Starobinsky model by a constant-roll model, in the sense that they have the same enhancement at the background level, i.e. the same $\Lambda^2=A_+^2/A_-^2=\epsilon_+/\epsilon_-$, we can express the $e$-folding number of the constant-roll USR stage by the integral of $\eta(N)$ in the Starobinsky model as:
\be\label{NUSR}
N_\text{USR}=-\frac16\ln\frac{\epsilon_-}{\epsilon_+}=-\frac16\int^{\epsilon=\epsilon_-}_{\epsilon=\epsilon_+}\eta(N)\dif N=\frac13\ln\Lambda,
\ee
where in the last step the second slow-roll parameter of the Starobinsky model, 
\be
\eta(N)=\frac{-6}{\displaystyle\frac{e^{3N}}{\Lambda-1} +1}
\label{etaa}
\ee
is used, which is derived in Appendix \ref{a:model}.

\begin{figure}
\begin{center}
\includegraphics[scale=0.7]{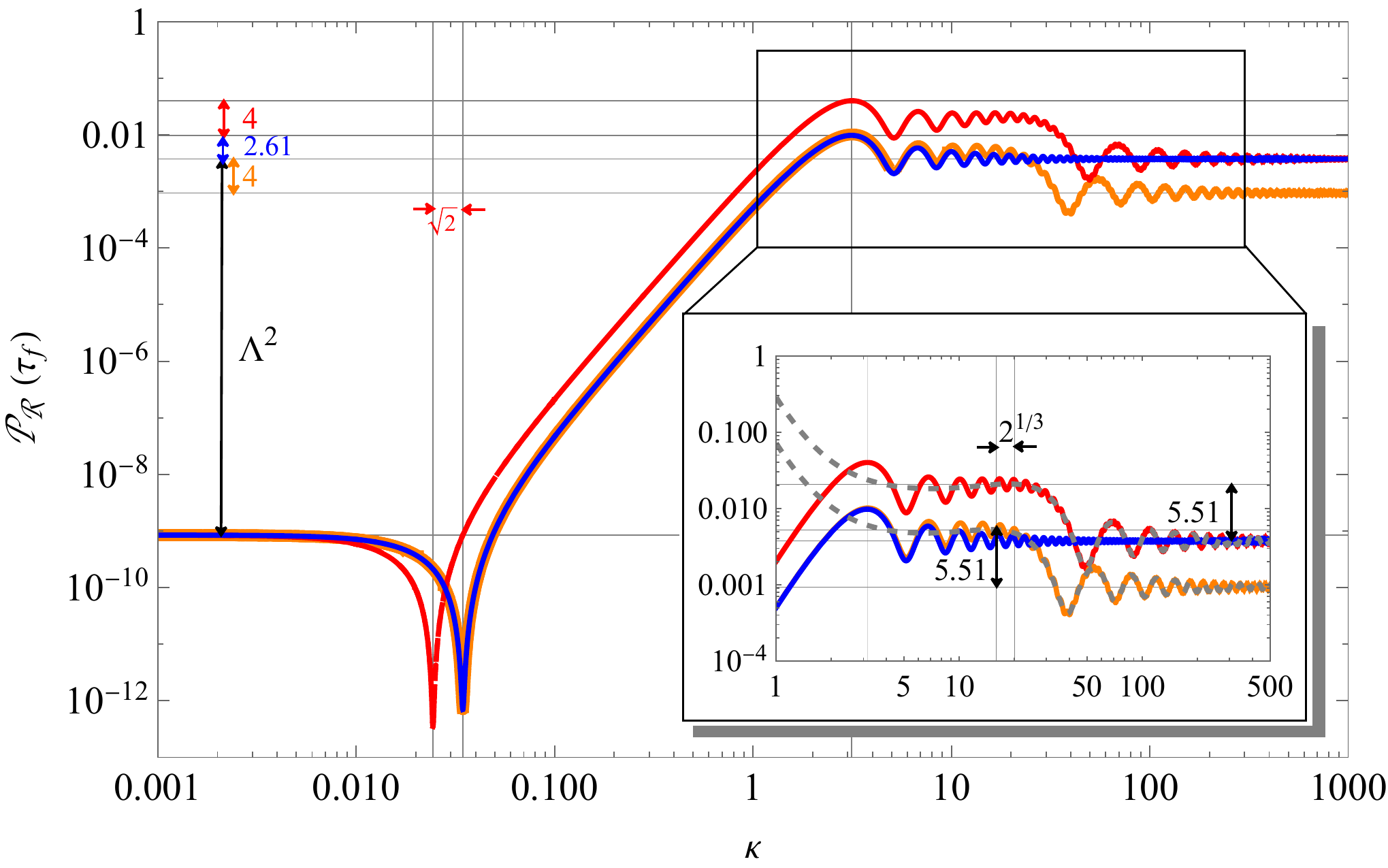}
\caption{The power spectrum $\mathcal{P_R}$ of the Starobinsky model (blue, given by (\ref{main})) and the constant-roll models, with the same IR and UV limit (red, given by (\ref{PR3}) with $N_\text{USR}=(1/3)\ln\Lambda$) and with the effective non-single-clock $e$-folding number (orange, given by (\ref{PR3}) with $N_\text{eff}^\text{(nsc)}=(1/3)\ln(\Lambda/2)$). We choose $\mathcal{P}_\mathcal{R}^\text{(IR)}=8.5\times10^{-10}$, $N_{\mathrm{USR}}=2.55$, $\Lambda=e^{3N_{\mathrm{USR}}}\approx2100$. For the red curve, the position of the dip is smaller by a factor of $\sqrt{2}$ than that of the Starobinsky model (see \eqref{sqrt2}), which induces a difference of 4 in their maxima (see \eqref{maxdiff}). The subfigure in right-bottom is a zoom-in of $\kappa>1$ region, which shows a difference of $5.51$ of the constant-roll plateau to its UV limit (\ref{551}). This is achieved by neglecting the high-frequency oscillations, shown by the gray dashed curve (see \eqref{exit}).}
\label{4times}
\end{center}
\end{figure}

The power spectrum in this model is \cite{Byrnes:2018txb}
\begin{align}
\mathcal{P}_{\mathcal{R}}(\kappa)
=&\mathcal{P}^{\mathrm{(IR)}}_{\mathcal{R}}\cdot \frac{\mathcal{G}_{0}+\left(e^{2 i \kappa} \mathcal{G}_{1}+e^{2 i\left(\kappa-\kappa_{1}\right)} \mathcal{G}_{2}+e^{2 i\left(\kappa-2 \kappa_{1}\right)} \mathcal{G}_{3}+e^{2 i \kappa_{1}} \mathcal{G}_{4}+\mathrm{c.c}\right)}{16 \kappa^{6}  \kappa_{1}^{6}}\cdot e^{6N_\mathrm{USR}},
\label{PR3}
\end{align}
where 
\be
\kappa_1\equiv-k\tau_1=\kappa e^{-N_{\mathrm{USR}}},
\ee
and $\mathcal{G}_{0,1,2,3,4}(\kappa)$ are some functions of $\kappa$ and $N_{\mathrm{USR}}$, written in Appendix \ref{a:constantroll}. This power spectrum is shown in Fig.\ref{4times}.  We immediately see that, besides the modulated oscillations by the abrupt slow-roll-to-USR transition at $\tau_0$, another low-frequency modulation appears. Besides, the enhancement of the power spectrum at background level (i.e. irrespective of the dynamics) is
\be\label{Pratio}
\left.\frac{\mathcal{P}_{\mathcal{R}}^{\mathrm{(UV)}}}{\mathcal{P}_{\mathcal{R}}^{\mathrm{(IR)}}}\right|_{\mathrm{CR}}=e^{6N_\mathrm{USR}},
\ee
which depends only on the $e$-folding number of USR phase.

As the constant-roll model has two instantaneous jumps of $\eta$, the excitation of negative-frequency modes will leave two sets of modulated oscillations accordingly with different frequencies. Besides, the maximum of power spectrum is again at its first crest, $\kappa_c\approx\pi$, which gives an extra parameter-independent enhancement factor:
\be
\left.\frac{\mathcal{P}_{\mathcal{R}}^{\text{(max)}}}{\mathcal{P}_{\mathcal{R}}^{\mathrm{(UV)}}}\right|_{\mathrm{CR}}\approx\frac{\mathcal{P}_{\mathcal{R}}\left( \kappa=\pi\right)}{\mathcal{P}_{\mathcal{R}}^{\mathrm{(UV)}}} \rightarrow 4+\frac{60}{\pi^{2}}+\frac{36}{\pi^{4}} =4\left( 1+\frac{15}{\pi^{2}}+\frac{9}{\pi^{4}} \right)=4\left. \frac{\mathcal{P}_{\mathcal{R}}^{\text {(max) }}}{\mathcal{P}_{\mathcal{R}}^{\mathrm{(UV)}}} \right|_{\mathrm{Star}}\approx 10.44,
\label{leading3}
\ee
where the subscript Star stands for Starobinsky model.
We see that when fixing the power spectra of both IR and UV slow-roll attractors, the peak in constant-roll model is 4 times larger than that in the Starobinsky model. This is also shown in Fig.\ref{4times}.

\begin{figure}
\begin{center}
\includegraphics[width=0.48\textwidth]{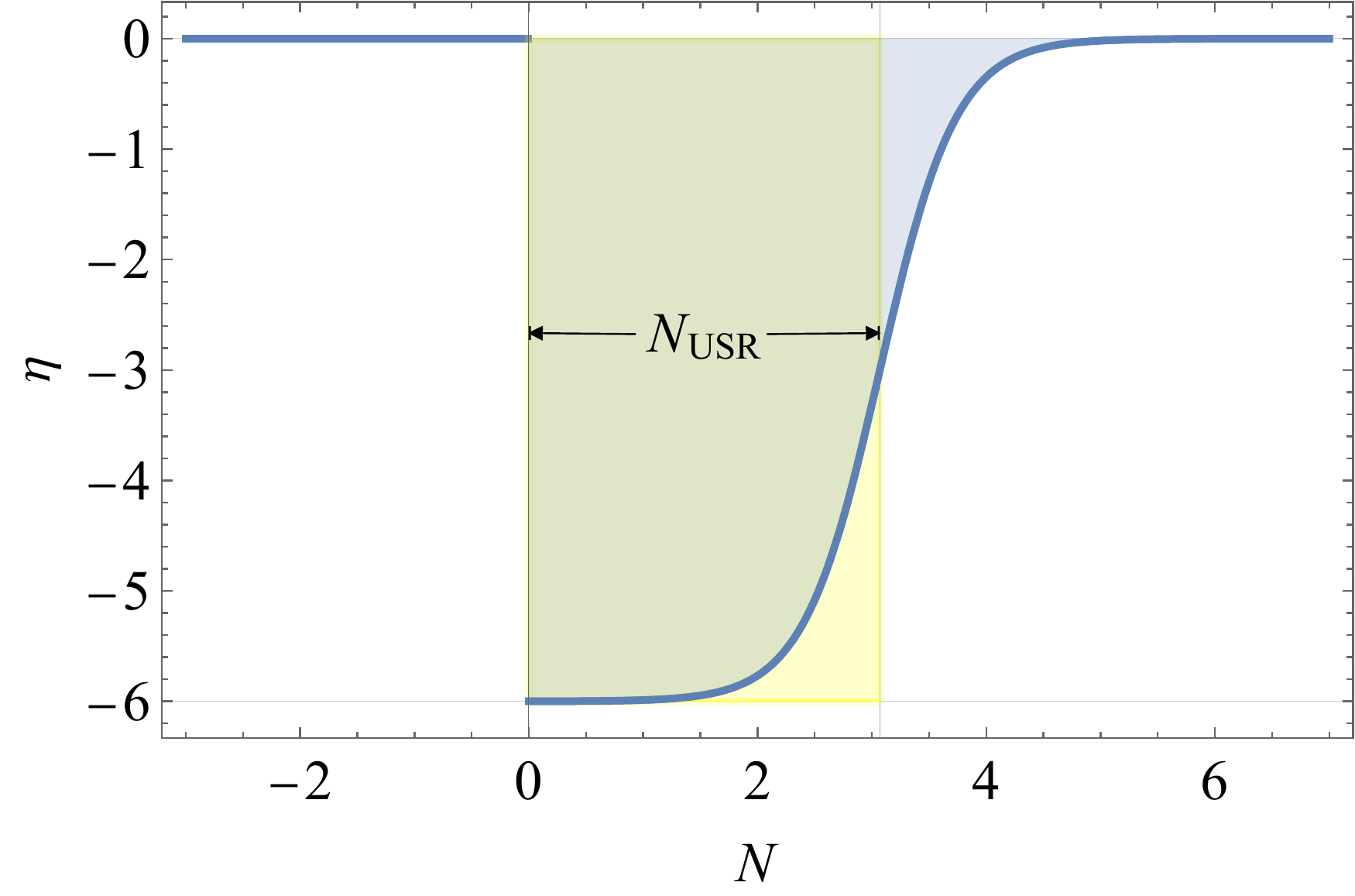}
\includegraphics[width=0.48\textwidth]{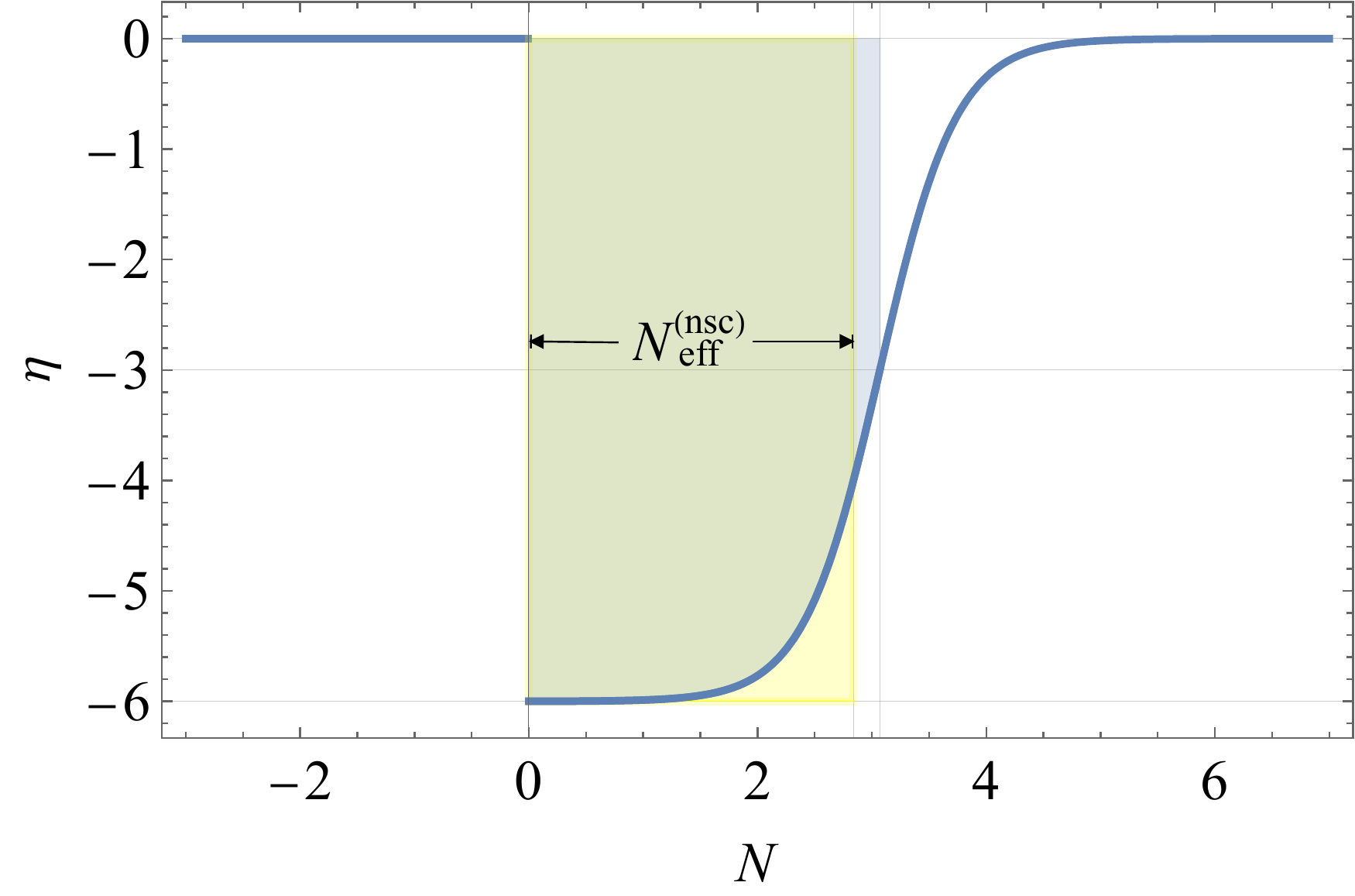}
\caption{$\eta(N)$ in Starobinsky model. Left: effective USR $e$-folding number $N_\text{USR}$, defined by \eqref{NUSR}. Right: effective non-single-clock $e$-folding number $N_\text{eff}^\text{(nsc)}$ which only includes contribution from non-single-clock stage ($\eta<-3$), \eqref{Neff}.}
\label{eta}
\end{center}
\end{figure}

This difference in the maximal value is interesting. Considering that the maximum are mainly contributed by the superhorizon evolution of the growing mode, as is shown in Fig.\ref{power spectrum}, we find that only the non-single-clock stage can contribute to the maximum. This inspires us to define another effective USR $e$-folding number which only takes the non-single-clock stage ($\eta<-3$) into account, 
\be
N_\text{eff}^\text{(nsc)}\equiv-\frac{1}{6}\int_{\eta<-3}\eta(N)\mathrm{d}N.
\ee
By the expression of $\eta(N)$ given in \eqref{etaa}, it is easy to find that the range for the non-single-clock phase in the Starobinsky model is $0<N<\frac13\ln(\Lambda-1)$, which, together with  \eqref{etaa}, gives the following effective $e$-folding number which only includes non-single-clock stage
\be
N_\text{eff}^\text{(nsc)}=-\frac{1}{6} \int_{0}^{\frac{1}{3}\ln(\Lambda-1)} \eta(N) \dif N=\frac{1}{3}\ln \left(\frac{\Lambda}{2}\right),
\label{Neff}
\ee
which is slightly smaller than $N_\text{USR}=(1/3)\ln\Lambda$, shown in Fig.\ref{eta}. Substituting \eqref{Neff} into \eqref{PR3}, we can immediately see that this gives the correct maxmimal value of the power spectrum, but a slightly suppresed UV plateau, as is shown in Fig.\ref{4times}.

We also find that by the virtue of the new $e$-folding number \eqref{Neff}, the position of the dip in constant-roll model found in Ref.~\cite{Byrnes:2018txb} (see (10) in \cite{Byrnes:2018txb})
 is consistent with the dip in Starobinsky model given in \eqref{dipposition}, as
\be
\kappa_\text{dip}=\sqrt{\frac54}e^{-\frac32N_\mathrm{eff}^\mathrm{(nsc)}}=\sqrt{\frac5{2\Lambda}}.
\ee
Therefore, the positions of the dips do not match if we define $N_\text{USR}$ as in \eqref{NUSR}, which in turn makes their maxima different. The difference of these two $e$-folding numbers is
\be
N_\mathrm{eff}^\mathrm{(nsc)}-N_{\mathrm{USR}}=-\frac{1}{3}\ln2,
\ee
which causes the dip in constant-roll model with $N_\text{USR}$ to appear earlier,
\be\label{sqrt2}
\frac{\kappa_\text{dip}(N_\text{eff}^\text{(nsc)})}{\kappa_\text{dip}(N_\text{USR})}=e^{ -\frac32\left( N_\mathrm{eff}^\mathrm{(nsc)}-N_{\mathrm{USR}} \right)}=\sqrt{2}.
\ee
As we already commented, the maximum always appears at the first crest of the modulated oscillation in these two models, i.e. $\kappa_c=\pi$. Therefore, due to the $k^4$-growth, the difference in the dips will render their maximal amplitudes different by a factor of 
\be\label{maxdiff}
\frac{\mathcal{P}_{\mathcal{R}}^{\mathrm{(max)}}\left(N_\text{USR}\right)}{\mathcal{P}_{\mathcal{R}}^{\mathrm{(max)}}\left(N_\text{eff}^\text{(nsc)}\right)}
=\left(\frac{\kappa_\text{dip}(N_\text{eff}^\text{(nsc)})}{\kappa_\text{dip}(N_\text{USR})}\right)^4=4.
\ee

The constant-roll model has to exit USR stage when $\eta$ goes from $-6$ to 0 instantaneously around $\kappa\sim e^{N_\mathrm{{USR}}}$, which will superimpose some new oscillations with lower frequency. Neglecting the remnant high-frequency oscillations generated at the first transition from slow-roll to USR, of which the amplitude  is now sufficiently small, we have the following simplified formula for the power spectrum with the new oscillations invoked by the second USR-to-slow-roll transition:
\be
\frac{\mathcal{P}^{\mathrm{(exit)}}_{\mathcal{R}}(\kappa)}{\mathcal{P}^{\mathrm{(UV)}}_{\mathcal{R}}}= \frac{\mathcal{G}_{0}+\left(\mathcal{G}_{4}e^{2 i\cdot \kappa_{1}} +\mathrm{c.c}\right)}{16\kappa^{6}\kappa_{1}^{6}}.
\label{exit}
\ee
As the period of the high-frequency oscillations is $\pi$, the period of these new oscillations is $\pi e^{N_\text{USR}}$. Similar to the maximum given by the high-frequency oscillations, we can also find a local maximum for the low-frequency oscillations of (\ref{exit}), after which the power spectrum begins to drop to its UV limit, 
\be\label{dropscale}
\kappa_{\mathrm{drop}}=\frac{\pi}{2} e^{N_\text{USR}}.
\ee
At this point, the power spectrum is 
\be
\frac{\mathcal{P}_{\mathcal{R}}^{\text {(drop)}}}{\mathcal{P}_{\mathcal{R}}^{\text {(UV)}}}=\lim _{N_{\mathrm{USR}} \to +\infty}\left.\frac{\mathcal{G}_{0}+\left(e^{2 i\kappa_{1}}   \mathcal{G}_{4}+\mathrm{c.c}\right)}{16   \kappa^{6}   \kappa_{1}^{6}}\right|_{\kappa=\frac{\pi}{2}   \exp(N_{\mathrm{USR}}), \kappa_{1}=\frac{\pi}{2}}=1+\frac{24}{\pi^2}+\frac{144}{\pi^4}+\frac{576}{\pi^{6}}\approx 5.51.
\label{551}
\ee

As a summary of this subsection, we find that the constant-roll model with appropriately defined effective USR $e$-folding number can mimic part of the power spectrum. If we fix both the IR and UV limits, the constant-roll model can produce an enhancement 4 times larger than the Starobinsky model, as its $e$-folding number of the USR phase is slightly larger. If we need the correct near-peak behavior, we should define $N_\text{eff}^\text{(nsc)}$ given by \eqref{Neff}, which only accounts the non-single-clock stage. This can give the amplitude of the peak correctly, at the price of generating a lower UV plateau.


\section{PBH Abundance and Induced GW}\label{s:pbh}
\subsection{Primordial Black Holes}
PBHs form via gravitational collapse, of which the mass function at their formation, $\beta(M)$, can be calculated, for instance by the Press-Schechter formalism \cite{Press:1973iz,Bond:1990iw}. As the probability distribution function (PDF) of the density perturbation $P[\delta\rho/\rho]$ has a tail that extends to large values, PBHs form when the density contrast on comoving hypersurfaces $\delta\equiv(\delta \rho/\rho)_c$ exceeds the critical density contrast $\delta_{\mathrm{cr}}$
~\cite{Zeldovich:1967lct,Hawking:1971ei,Carr:1974nx,Meszaros:1974tb,Carr:1975qj,Khlopov:1985jw}, determined by numerical simulations~\cite{Musco:2004ak,Musco:2008hv,Musco:2012au,Musco:2018rwt,Germani:2018jgr,Germani:2019zez,Escriva:2019nsa,Escriva:2019phb,Escriva:2020tak}. In this paper we choose $\delta_{\mathrm{c r}}=0.41$ for simplicity~\cite{Harada:2013epa}. In the Press-Schechter formalism, $\beta$ is calculated by integrating the PDF of the density contrast $P[\delta]$ from $\delta_{\mathrm{cr}}$,
\be
\beta=2\int_{\delta_{\mathrm{cr}}} P[\delta] \dif \delta,
\ee
where the PDF is usually Gaussian:
\be
P[\delta]=\frac{1}{\sqrt{2 \pi \sigma_{\delta}^{2}}} \exp \left(-\frac{\delta^{2}}{2 \sigma_{\delta}^{2}}\right).
\ee
$\sigma_\delta^2$ is the smoothed variance of the density contrast which will be defined soon.
 
The density contrast $\delta(\tau, \mathbf{x})$ in real space can be expanded in the momentum space with Fourier mode $\delta_{\mathbf{k}}(\tau)$, such that its power spectrum can be written as
\be\label{definePR}
\mathcal{P}_\delta(k)=\frac{k^3}{2\pi^2}|\delta_{\mathbf{k}}|^2.
\ee
In order to  smear out fluctuations smaller than a comoving scale $R$, one should define a smoothed density contrast $\delta(\mathbf{x} ; R)$ by convolving the real-space contrast with a window function $W$, 
\be
\delta(\mathbf{x} ; R)=\int \dif^{3} \mathbf{x}'~W\left(\left|\mathbf{x}-\mathbf{x}^{\prime}\right| ; R\right) \delta\left(\mathbf{x}^{\prime}\right)=\int \frac{\dif^{3} \mathbf{k}}{(2 \pi)^{3 / 2}} \widetilde{W}(k ; R) \delta_{\mathbf{k}} e^{i \mathbf{k} \cdot \mathbf{x}},
\label{deltatauxR}
\ee
where $\widetilde{W}(k;R)$ is the Fourier mode of the window function. The variance smoothed on comoving scale $R$ is defined as the two-point correlation function of $\delta(\mathbf{x} ; R)$ at the same point, 
\be
\sigma^2_{\delta}(R) \equiv \lim\limits_{\mathbf{y}\to \mathbf{x}} \langle\delta(\mathbf{x} ; R) \delta(\mathbf{y} ; R)\rangle=\int \frac{\dif k}{k} \widetilde{W}^{2}(k ; R) \mathcal{P}_{\delta}(k).
\label{sigmadef}
\ee
For simplicity, we only consider the standard thermal history with a radiation dominated era, hence the power spectrum of $\delta$ can be expressed by that of the comoving curvature perturbation as
\be
\mathcal{P}_{\delta}(k)=\frac{16}{81}\left(\frac{k}{H a}\right)^{4} \mathcal{P}_{\mathcal{R}}(k).
\label{deltaR}
\ee
To account for the PBH formation, we choose the smoothing scale of the window function $R$ to be the comoving horizon $1/(Ha)$, and substitute (\ref{deltaR}) into (\ref{sigmadef}), 
 \be\label{variancedelta}
\sigma_{\delta}^{2}(R)=\frac{16}{81} \int \frac{\dif k}{k}(k R)^{4} \mathcal{P}_{\mathcal{R}}(k) \widetilde{W}^2(k ; R).
\ee
In this paper we choose the real-space Gaussian window function, whose Fourier mode is
\be
\widetilde{W}(k ; R)=\exp \left(-\frac{k^{2} R^{2}}{2}\right).
\label{GaussianWindow}
\ee
 For discussion on the uncertainties from different choices of window functions, see Refs.\cite{Young:2019osy,Tokeshi:2020tjq,Gow:2020cou}. Having these conditions in mind, we can calculate the PBH mass function at its formation
\be
\beta(M)=2 \frac{\gamma}{\sqrt{2 \pi \sigma_{\delta}^{2}(M)}} \int_{\delta_{\mathrm{cr}}}^{\infty} \exp \left(-\frac{\delta^{2}}{\sigma_{\delta}^{2}(M)}\right) \dif \delta\approx \frac{2 \gamma}{\sqrt{2 \pi}   \nu(M)}   \exp \left(-\frac{\nu^{2}(M)}{2}\right),
\ee
where $\nu(M)\equiv\delta_\mathrm{cr}/\sigma_\delta(M)$, and $\gamma\sim0.2$ is the fraction of  horizon mass that collapses into a PBH once $\delta>\delta_\mathrm{cr}$. Here we already change the variable from the comoving horizon $R$ to the PBH mass $M$, which are connected by $M=\gamma Ra/(16G)$. 

\begin{figure}
\begin{center}
\subfigure[$\Lambda=4207$, $ M_c=1.72\times 10^{17}\mathrm{g}$]{\includegraphics[width=0.49\textwidth]{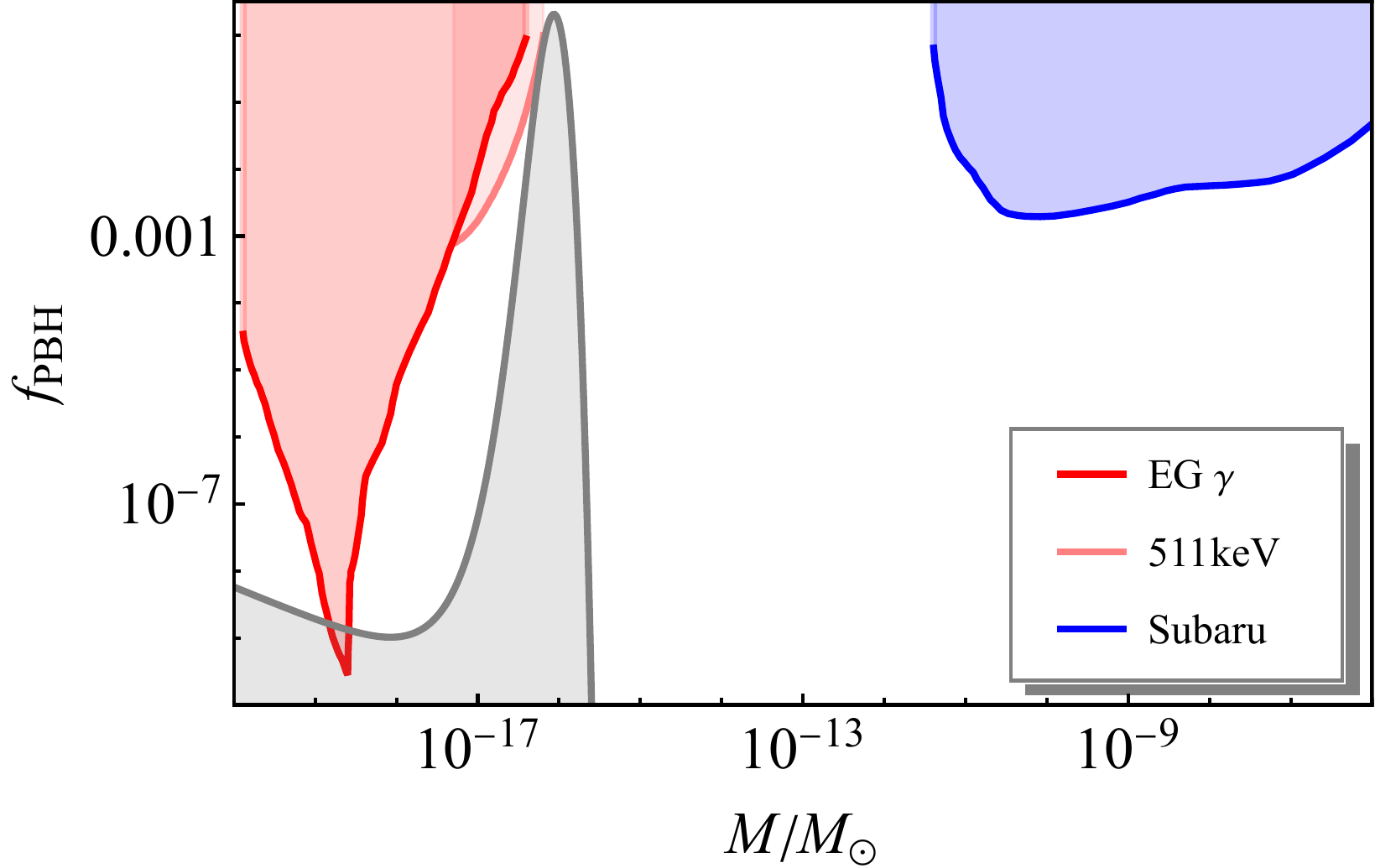}}
\subfigure[$\Lambda=4574$, $ M_c=5.85\times 10^{21}\mathrm{g}$]{\includegraphics[width=0.49\textwidth]{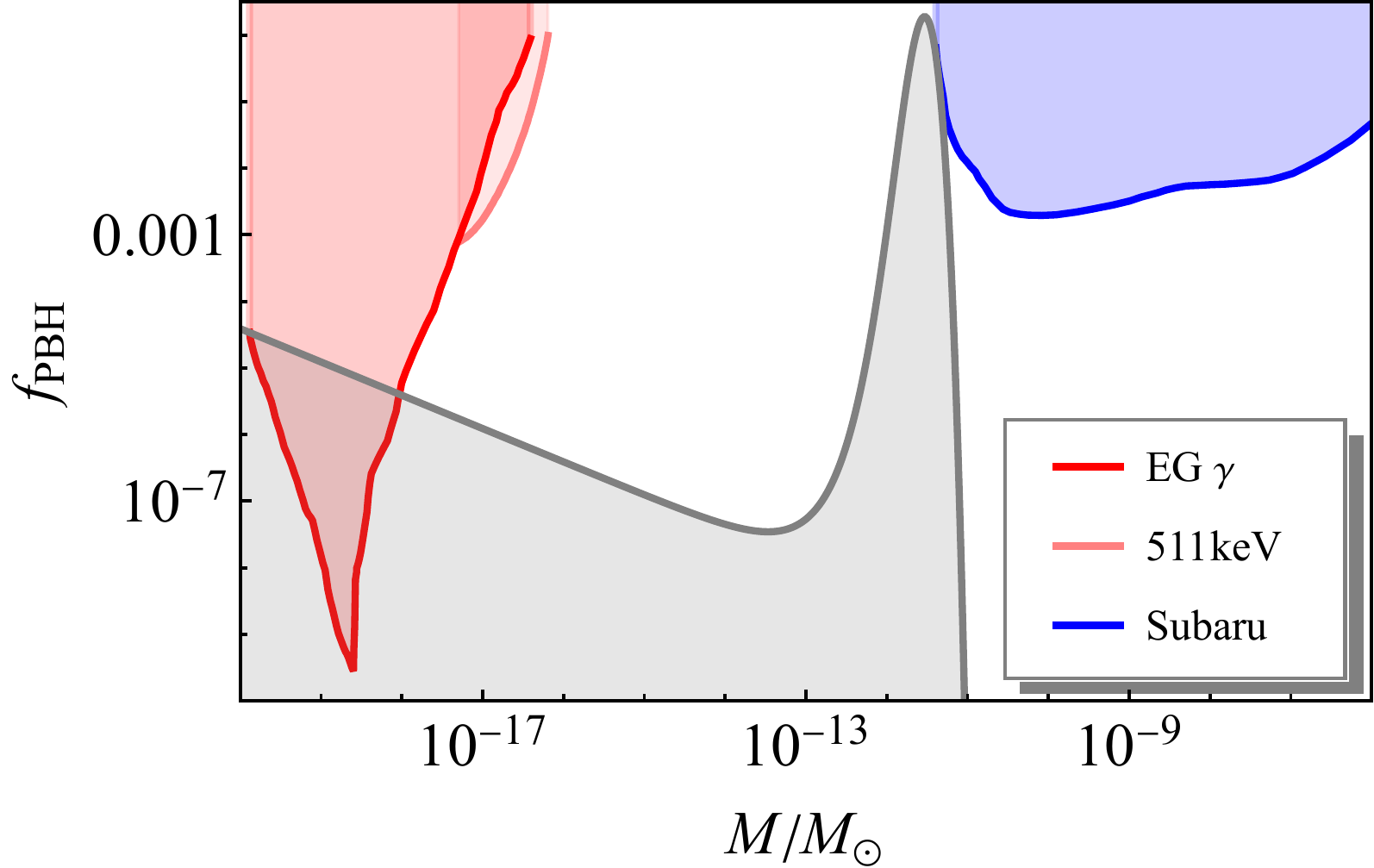}}
\caption{The PBH mass function $f_\text{PBH}(M)$ in the Starobinsky model is drawn together with the observational constraints from extra-galactic gamma ray \cite{Carr:2009jm,Carr:2020gox}, the 511 keV line from the galactic center \cite{DeRocco:2019fjq,Laha:2019ssq,Dasgupta:2019cae,Ray:2021mxu}, and microlensing by Subaru HSC~\cite{Niikura:2017zjd,Sugiyama:2019dgt}. In the left and right panel, the peak mass is $ M_c=1.72\times10^{17}~\text{g}$ and $ M_c=5.85\times10^{21}~\text{g}$, respectively. (Solar mass $M_{\odot}=1.99\times 10^{33}~\text{g}$.) Note that according to the different redshifts due to the different moments of formation during the radiation-dominated era, even for a fixed abundance of PBH, the maximum of their mass functions, described solely by $\Lambda$, differs slightly. }
\label{fPBH}
\end{center}
\end{figure}

After redshifted in the radiation dominated era, the PBH abundance today, defined as the energy density of PBHs normalized by that of the cold dark matter, can be written as \cite{Carr:2016drx} 
\be
f_{\mathrm{PBH}}(M)=1.65 \times 10^{8}\left(\frac{\gamma}{0.2}\right)^{1 / 2}\left(\frac{g_{*}}{106.75}\right)^{-1 / 4}\left(\frac{h}{0.68}\right)^{-2} \beta(M) \left(\frac{M}{M_{\odot}}\right)^{-1/2},
\ee
$g_{*}$ is the effective relativistic degree of freedom at the formation, and $h\equiv H_0/(100~\text{km/s/Mpc})$ is the normalized Hubble constant. 
Suppose that PBH constitutes all the dark matter, we have $\int f_\mathrm{PBH}(M)\dif\ln M=1$, which can be used to determine the amplitude of the peak in the primordial power spectrum of the comoving curvature perturbation. In the Starobinsky model we considered, it is $\Lambda$ to be determined by the condition of $\int f_\mathrm{PBH}(M)\dif\ln M=1$, as the power spectrum on large scales is fixed. According to the current observational constraints, the only possible mass interval where PBH can be all the dark matter is the asteroid-mass window, $1.72\times 10^{17}~\mathrm{g}$ to $5.85\times 10^{21}~\mathrm{g}$, which corresponds to $\Lambda=4207\sim4574$ if we fix $\mathcal{P}_\mathcal{R}^\text{(IR)}=8.5\times10^{-10}$. The PBH mass functions with these parameters are shown in Fig.\ref{fPBH}.

\begin{figure}
\begin{center}
\subfigure[$N_{\mathrm{USR}}=2.55$, $ M_c=1.80\times 10^{17}\mathrm{g}$]{\includegraphics[width=0.49\textwidth]{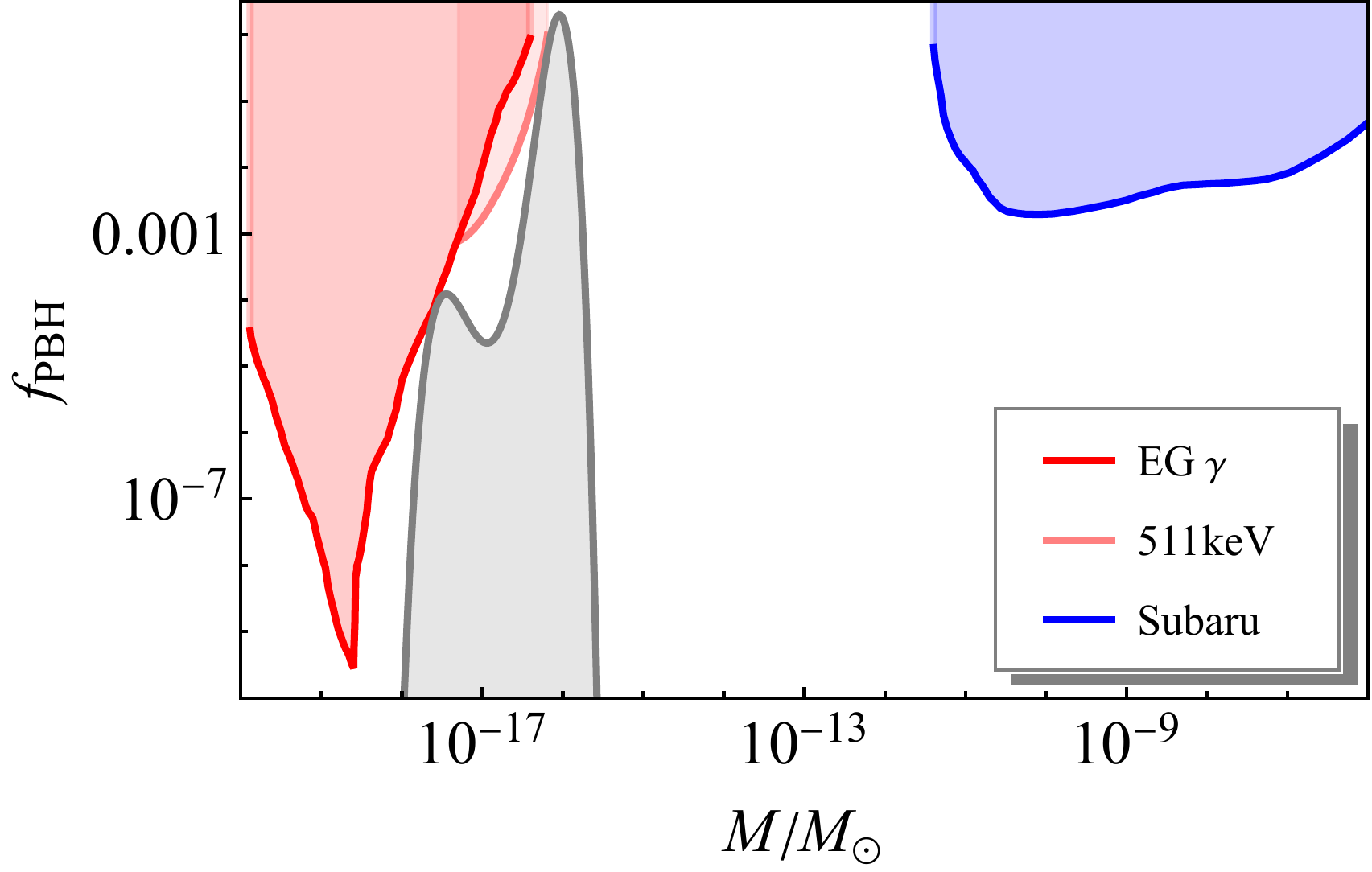}}
\subfigure[$N_{\mathrm{USR}}=2.57$, $ M_c=5.75\times 10^{21}\mathrm{g}$]{\includegraphics[width=0.49\textwidth]{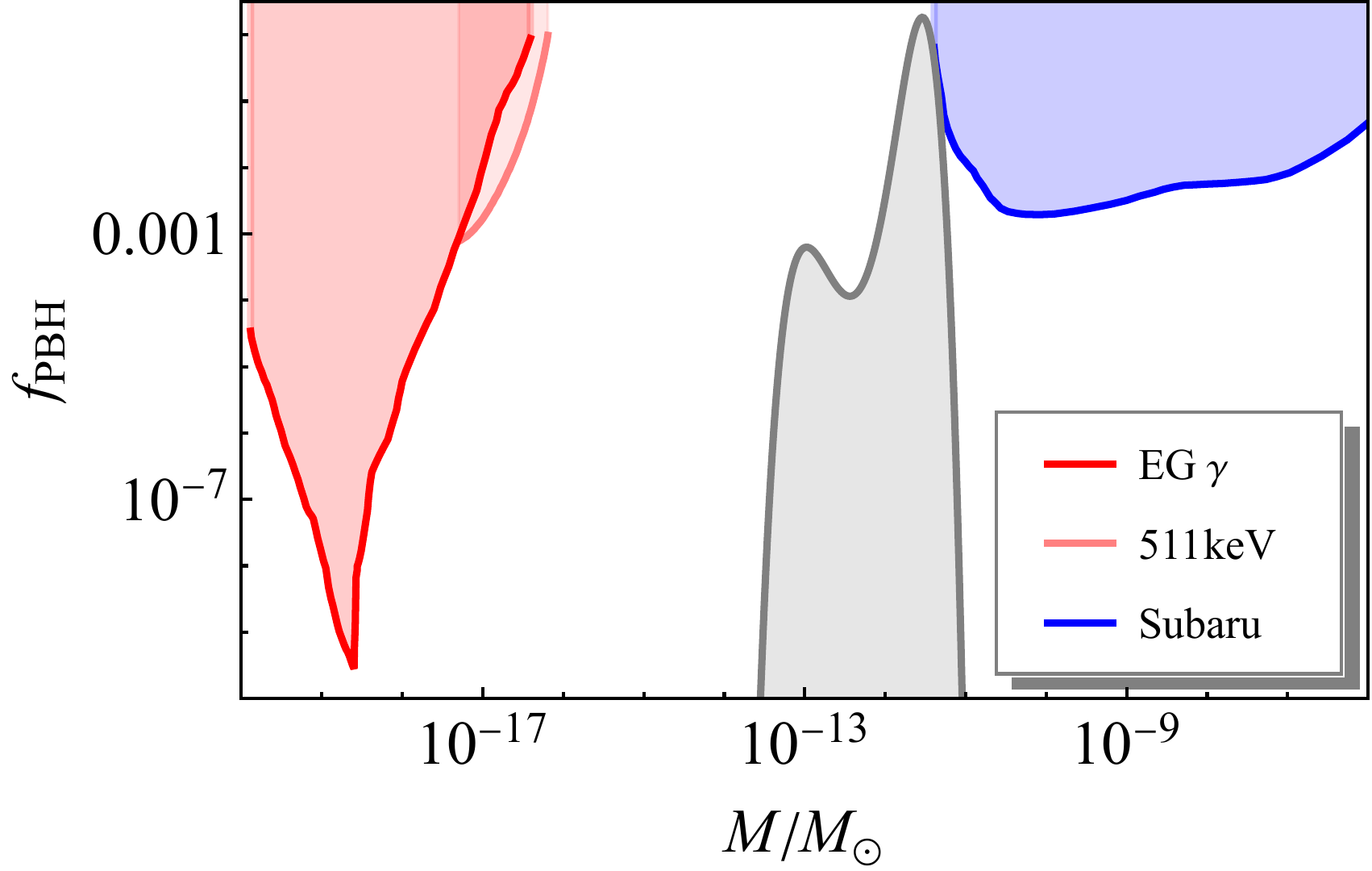}}
\caption{
The PBH mass function $f_\text{PBH}$ in the constant-roll model is drawn together with the same observational constraints as Fig.\ref{fPBH}.
Left and right panels have the peak mass $ M_c=1.80\times10^{17}~\text{g}$ and $ M_c=5.75\times10^{21}~\text{g}$, respectively.}
\label{fPBH3}
\end{center}
\end{figure}

It is obvious from Fig.\ref{fPBH} that when PBH constitutes all the dark matter, the PBH abundance derived from the Starobinsky model already contradicts with the constraint from the extra-galactic gamma ray. This is because in the Starobinsky model, the maximum from modulated oscillation is only $\sim2.61$ times larger than the plateau, as is shown in \eqref{nextleading}, which is transferred to ``only'' 8 orders of magnitude larger in the mass function. The null detection of extra-galactic gamma ray as the remnant of the Hawking radiation has constrained the tiny PBHs $\sim10^{14}~\text{g}$ to be smaller than $10^{-10}$, then the aforementioned peak of $\sim\mathcal{O}(1)$ will inevitably generate too many tiny PBHs. Therefore, a realistic model should suppress the plateau further. The constant-roll model we considered in Section \ref{s:constantroll} can work well, as its maximum is 4 times larger than that of the Starobinsky model, which is $4\times2.61\sim 10$ times larger than the UV plateau, circumventing the gamma-ray constraints when PBHs are all the dark matter, as is shown in Fig.\ref{fPBH3}. 
In the constant-roll model there are two peaks in the mass function. The main peak comes from the first crest of the oscillations. The power law in the mass function derived from the redshift, $M^{-1/2}$ , causes an increase towards smaller PBHs. So the UV edge of the plateau in the power spectrum of $\mathcal{R}$ gives a lower peak in the mass function of PBH. 

Similar to the previous discussion for the Starobinsky model, we can calculate its corresponding USR $e$-folding number $N_{\mathrm{USR}}$. The possible mass window for PBH to be all the dark matter is slightly smaller than that of the Starobinsky model, namely $1.80\times 10^{17}~\mathrm{g}$ to $5.75\times 10^{21}~\mathrm{g}$. The corresponding $N_{\mathrm{USR}}$ are from $2.55$ to $2.57$ if we fix $\mathcal{P}_\mathcal{R}^\text{(IR)}=8.5\times10^{-10}$. 

\subsection{Induced Gravitational Waves}
Now we turn to the stochastic GW background induced by the scalar perturbation via the nonlinear scalar-scalar-tensor type interaction, following \cite{Pi:2020otn,Domenech:2021ztg}. We skip the details and directly write down the formula for the induced GW spectrum we observe today:
\begin{align}
\Omega_\text{GW}(f)h^2
&=1.6\times10^{-5}\left(\frac{g_{*s}(\tau_k)}{106.75}\right)^{-1/3}\left(\frac{\Omega_{r}h^2}{4.1\times10^{-5}}\right)\Omega_\text{GW,eq}(f),
\end{align}
where $g_{*s}(\tau_k)$ is the effective relativistic degree of freedom for the entropy density, $\Omega_{r}$ is the cosmological parameter for radiation today, and $\Omega_\text{GW,eq}(f)$ is the GW spectrum at the matter-radiation equality, given by
\begin{align}
\label{Omega0}
\Omega_\text{GW,eq}(k)
&=3\int^\infty_0dv\int^{1+v}_{|1-v|}du\frac{\mathcal{T}(u,v)}{u^2v^2}\mathcal{P}_\mathcal{R}(vk)\mathcal{P}_\mathcal{R}(uk),\\\nn
\mathcal{T}(u,v)&=\frac14\left[\frac{4v^2-(1+v^2-u^2)^2}{4uv}\right]^2\left(\frac{u^2+v^2-3}{2uv}\right)^4\\
&\cdot\left[\left(\ln\left|\frac{3-(u+v)^2}{3-(u-v)^2}\right|-\frac{4uv}{u^2+v^2-3}\right)^2+\pi^2\Theta\left(u+v-\sqrt3\right) \right].
\end{align}
The comoving wavenumber $k$ is connected to the frequency $f$ by $f=k/(2\pi a_0)\approx1.5\times10^{-9}(k/1~\text{pc}^{-1})~\text{Hz}$.

\begin{figure}
\begin{minipage}{0.48\textwidth}
\begin{center}
\includegraphics[width=1\textwidth]{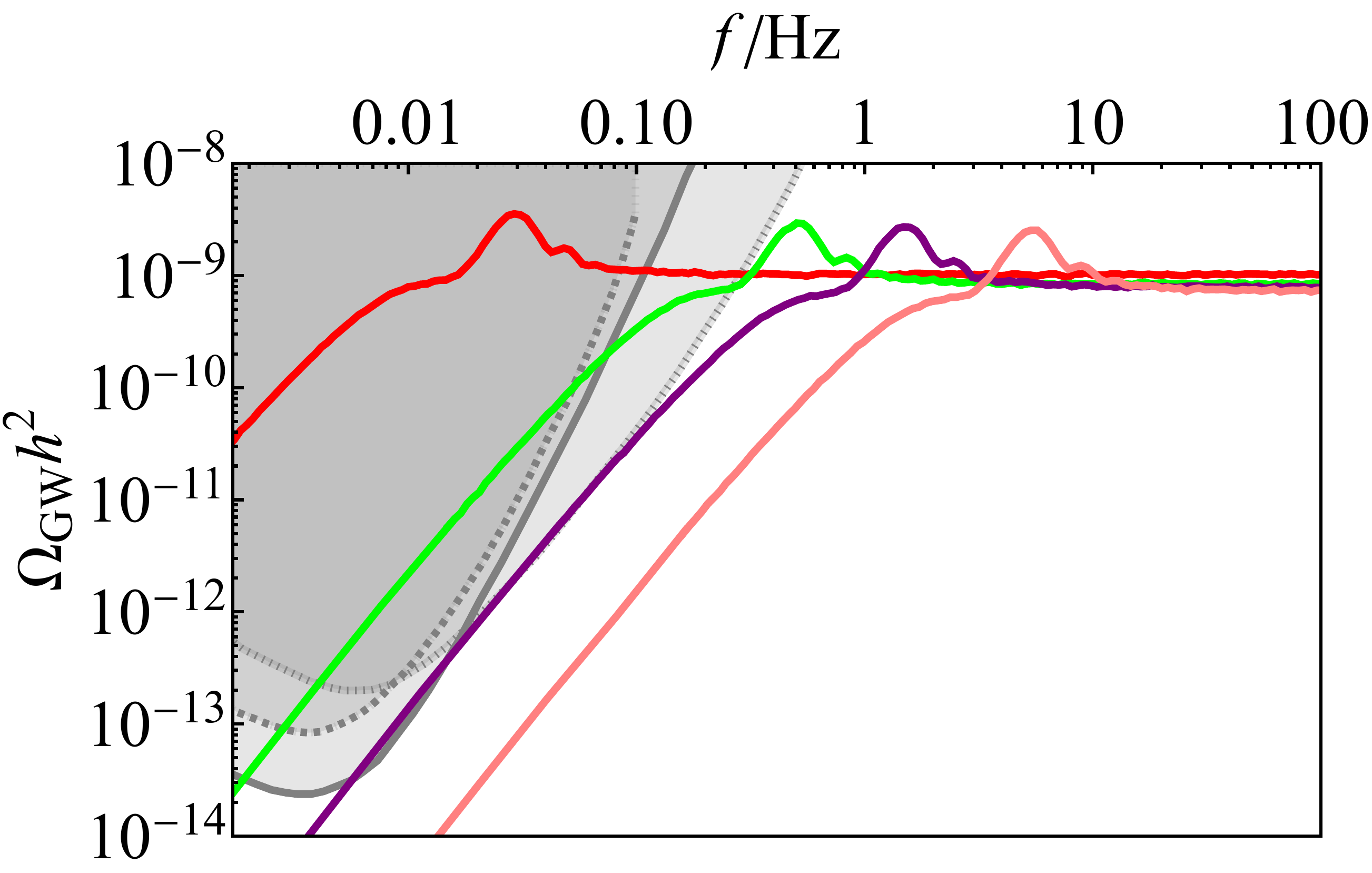}
\end{center}
\end{minipage}
\begin{minipage}{0.48\textwidth}
\begin{center}
\includegraphics[width=1\textwidth]{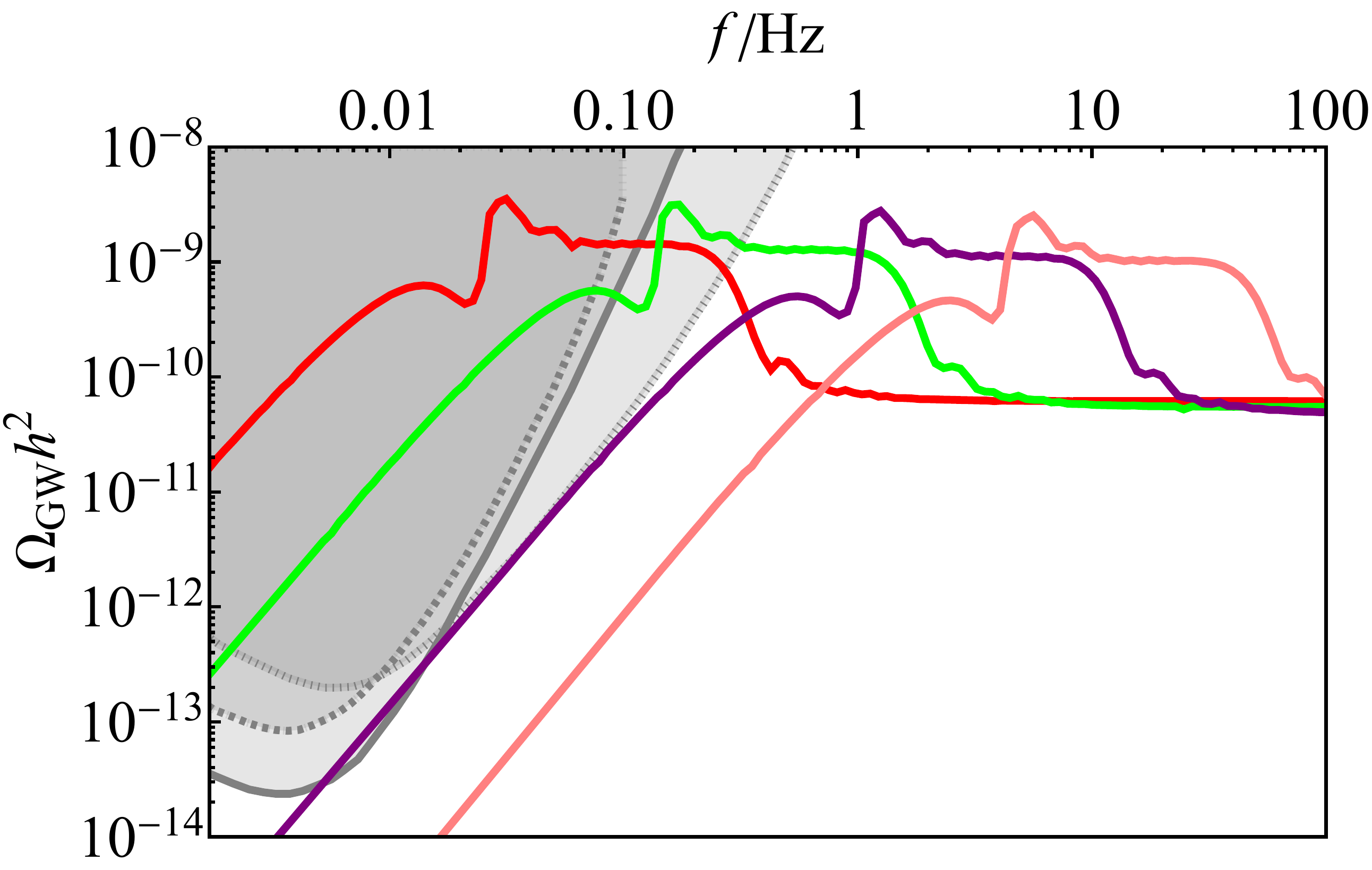}
\end{center}
\end{minipage}
\\
\vspace{2ex}
\begin{minipage}{0.48\textwidth}
\begin{center}
\setlength{\tabcolsep}{0.7em}{
 \begin{tabular}{c c c c} 
 \hline
curves & $f_p/$Hz & $ M_c$/g  \\
 \hline
red & $2.93\times 10^{-2}$ & $5.85\times 10^{21}$   \\ 
green & $5.17\times 10^{-1}$ & $1.95\times 10^{19}$   \\
purple & $1.55\times 10^{0}$ & $2.25\times 10^{18}$   \\
pink & $5.56\times 10^{0}$ & $1.72\times 10^{17}$   \\
 \hline
\end{tabular}}
\end{center}
\end{minipage}
\begin{minipage}{0.48\textwidth}
\begin{center}
\setlength{\tabcolsep}{0.7em}{
 \begin{tabular}{c c c c} 
  \hline
curves & $f_p$/Hz & $ M_c$/g  \\
 \hline
red & $~3.11\times 10^{-2}~$ & $~5.75\times 10^{21}~$   \\ 
green & $1.66\times 10^{-1}$ & $1.92\times 10^{20}$   \\
purple & $1.25\times 10^{0}$ & $3.65\times 10^{18}$   \\
pink & $5.62\times 10^{0}$ & $1.80\times 10^{17}$   \\
\hline
\end{tabular}}
\end{center}
\end{minipage}
\vspace{1ex}
\caption{The GW spectrum $\Omega_{\mathrm{GW}}\left(f\right)$ with different peak frequencies $f_p$ in the Starobinsky model (left) and constant-roll model (right), together with the sensitivity curves of LISA \cite{Bartolo:2016ami} (solid), Taiji~\cite{Wang:2021njt} (dashed), and TianQin~\cite{Liang:2021bde} (dotted), provided PBHs with peak mass $M_c$ to be all the dark matter.}\label{IGW}
\end{figure}

We show $\Omega_\mathrm{GW}\left(f\right)h^2$ of the Starobinsky model and constant-roll model with the sensitivity curves of LISA \cite{Bartolo:2016ami}, Taiji~\cite{Wang:2021njt}, and TianQin~\cite{Liang:2021bde} in Fig.\ref{IGW}.\footnote{As we have shown, the global Starobinsky model is excluded by the constraints on extragalactic gamma ray. So a realistic model might include another step-down Starobinsky model to suppress the power on smaller scales, which generates a suppression in the spectrum of the induced GWs. This is not reflected in Fig.\ref{IGW} as this suppression is model dependent, and difficult to be detected in the near future.} To identify the origin of such GW signals, we should distinguish them from those in other models by their characteristic features. It is obvious that the low-frequency part of $\Omega_\text{GW}$ grows as a power-law of $f^3$, which is the universal infrared scaling of GW spectrum~\cite{Cai:2019cdl}. This scaling can be used to probe the thermal history of the universe~\cite{Domenech:2019quo,Domenech:2020kqm,Hook:2020phx,Witkowski:2021raz,Caldwell:2022qsj,Brzeminski:2022haa}, but unable to distinguish models. The high-frequency part is a scale-invariant plateau, reflecting the final recovery of the slow-roll inflation \footnote{On the attractor solution for a constant $\eta$, $\mathcal{P_R}$ goes as a power law of $\propto k^{3-|3+\eta|}$~\cite{Byrnes:2018txb}. So only for $\eta=0$ and $-6$, the power spectra have scale-invariant plateaus. For the constant-roll model we considered, $\eta$ goes as $0\to-6\to0$, so three plateaus appears as expected.}. However, such a scale-invariant GW spectrum could appear in many different mechanisms, thus can not be used to distinguish them either \cite{Carr:2019kxo,Ellis:2020ena,Blasi:2020mfx,DeLuca:2020agl,Domenech:2020ers,Kusenko:2020pcg,Balaji:2022dbi}. 

Hence we should focus on the characteristic features of the modulated oscillations, especially the shape around the first peak, which originates from the modulated oscillation. As the period is $\pi$, this peak is narrow, since its dimensionless width is $\Delta\sim(2\pi)^{-1}\sim0.16$, smaller than the critical width of $0.4$, given in Ref.\cite{Pi:2020otn}. In the GW spectrum, an obvious dip before the peak should have appeared due to the destructive interference in the convolution of the scalar perturbations, 
if the height of the peak were much larger than that of the other parts \cite{Balaji:2022dbi}. However, in the models we studied, the peak is not high enough, so the dip is blurred by the contribution from the UV plateau. Furthermore, as the relative amplitude of the peak in the Starobinsky model ($\sim2.61$, see Eq.\eqref{nextleading}) is smaller than that in the constant-roll model ($\sim10$, see Eq.\eqref{leading3}), the dip is more obscure in the Starobinsky model, as is shown in Fig.\ref{IGW}. Nevertheless, such a shoulder/dip as well as the peak-to-plateau ratio $\sim2.61^2$ or $\sim10^2$ are the characteristic features in the GW spectrum, implying the appearance of modulated oscillations from an abrupt slow-roll-to-USR transition, as the period as well as the relative amplitude of the modulated oscillations do not depend on the parameters. The shape of such features are quite different from other GW spectra which contain modulated oscillations, like the multi-field inflation \cite{Fumagalli:2020adf,Fumagalli:2020nvq,Fumagalli:2021cel}, excited states \cite{Fumagalli:2021mpc}, resonance peaks \cite{Cai:2019amo}, first-order phase transition \cite{An:2020fff,An:2022cce}, etc. Therefore, these features can be seen as a smoking gun for an abrupt slow-roll-to-USR transition, which could be probed by the space-borne interferometers in the next decades~\cite{Caprini:2019pxz,Fumagalli:2021dtd}.


We would like to comment that not all the GW spectrum associated with PBHs as all the dark matter can be detected by LISA/Taiji/TianQin. Because the sensitivity for such interferometers at decihertz is poor, a window of small asteroid-mass PBH $\lesssim10^{18}~\text{g}$ are undetectable by the next-generation interferometers, as is shown clearly in Fig.\ref{IGW}. However, they can be probed in the near future by either the MeV gamma ray \cite{Ray:2021mxu} or by the third generation space-borne interferometers like DECIGO~\cite{Kawamura:2011zz} and BBO~\cite{Crowder:2005nr,Corbin:2005ny,Baker:2019pnp}.


\section{Conclusion and discussion}\label{s:conclusion}

We study the scalar power spectrum in the Starobinsky model. The analytical discussion proceeds from direct matching the vacuums and the gradient expansion method. Both the methods coincide with the numerical result very well. While the main enhancement is provided by the suppression of the slow-roll parameter $\epsilon$, the modulated oscillations caused by the transient excitation of the negative-frequency modes can generate an extra enhancement which gives the global maximum at its first crest. The position is $\pi$ times the transition scale, while the extra enhancement factor is $1+15/\pi^2+9/\pi^4\approx2.61$. Both of them are independent of the parameters of the model, if $\Lambda\gg1$. 

We find that the constant-roll model can mimic the Starobinsky model quite well, as long as we appropriately define the effective USR $e$-folding number:
\be\label{def:Neff}
N_\mathrm{USR}=-\frac16\int \eta(N)\dif N, \quad
\text{or}\quad
N_\mathrm{eff}^\mathrm{(nsc)}=-\frac16\int_{\eta<-3}\eta(N)\dif N.
\ee 
When fixing the IR power spectrum, the former definition goes to the correct UV limit, but gives a larger maximum. However, the latter definition, which only includes the non-single-clock stage of $\eta<-3$, generates the correct maximum, at the price of suppressing the UV plateau. As the instant transition of $\eta$ may be unphysical \cite{Cole:2022xqc}, a realistic model usually has a varying $\eta(N)$. In such a case, we conjecture that $N_\mathrm{eff}^\mathrm{(nsc)}$ can be used to estimate the peak of the power spectrum, when it is mainly contributed by the non-single-clock stage. This will be left for future work.


The dip of the power spectrum, which locates at $\sqrt{5/(2\Lambda)}$ or $\sqrt{5/4}\exp\left(-\frac32N_\mathrm{eff}^\mathrm{(nsc)}\right)$ multiplying the transition wavenumber, is another important feature of the power spectrum. Unfortunately the amplitude of the dip is too small to be detected directly. But recent developments show that it is possible to probe this dip on CMB $\mu$-distortions as the dip induces large non-Gaussianities \cite{Ozsoy:2021pws} via the Maldacena's consistency relation \cite{Maldacena:2002vr}, which still holds around the dip. The peak on such a large scale can generate supermassive PBHs as seeds for galaxy or structure formation \cite{Nakama:2017xvq,Carr:2020erq}, which is beyond the scope of our paper.

When applying our result to the PBH mass function, we find that the ``global'' Starobinsky model predicts too many tiny PBHs of $M_c\sim10^{14}~\text{g}$, which contradict the constraints from extra-galactic gamma ray, as expected. This is because the extra enhancement of $\sim2.61$ is not large enough. Adding another step-down Starobinsky potential after when the inflaton is already back to the slow-roll attractor can effectively suppress the small-scale power spectrum, which is the prototype of the smoothed near-inflection point inflation well studied recently. Additionally, the power spectrum in the constant-roll model has a further drop, which can also pass the gamma ray constraints.

The spectrum of the induced GWs has characteristic features due to the modulated oscillations. Although the modulated oscillations only leave insignificant imprints in $\Omega_\text{GW}$, the characteristic features of an infrared shoulder/dip, a narrow peak, and a flat UV plateau with a specific peak-to-plateau ratio are universal for all the models with an abrupt slow-roll-to-USR transition. Therefore, these features on the stochastic GW spectrum, which might be detected in the space-borne interferometers in the next decade, can be used to verify such a scenario. 

In this paper we consider only Gaussian perturbations, and neglect the back reaction of the quantum fluctuations. As we know, the PBH abundance ~\cite{Young:2013oia,Yoo:2019pma,Atal:2019cdz,Biagetti:2021eep} as well as its clustering  \cite{Desjacques:2018wuu,Suyama:2019cst,Franciolini:2018vbk} are very sensitive to non-Gaussianity, which can also leave characteristic features on the induced GWs \cite{Garcia-Bellido:2016dkw,Nakama:2016gzw,Cai:2018dig,Unal:2018yaa,Adshead:2021hnm,Garcia-Saenz:2022tzu}. However, it is shown that the non-Gaussianity is always negligible unless the exit of USR phase is abrupt \cite{Cai:2018dkf,Passaglia:2018ixg}. This means for Starobinsky model, as the inflaton field rolls back to slow-roll attractor slowly, the non-Gaussianity is washed out. However, in both the step-down Starobinsky model and the constant-roll model (when $\eta$ goes from $-6$ to $\sim0$ suddenly), non-Gaussianity could be large. It depends on the field velocity at the transition, which can be as large as $f_\text{NL}=5/2$ when this velocity is much smaller than that of the final slow-roll phase. This might further change the amplitude and shape of the scalar power spectrum, which breaks the robustness of the shape, yet makes it possible to probe the duration of the transition. We will leave this issue for future work. On the other hand, when considering quantum diffusion, as is recently studied in Ref.\cite{Pattison:2021oen}, the Gaussian PDF is a good approximation in the Starobinsky model as long as the potential is not too flat, i.e. $A_-\mpl/V_0\gtrsim H_0/\mpl$. This gives $A_-\gtrsim H_0^3$, a condition well satisfied in our parameter choice, thus we do not have to worry about the quantum diffusion. 

\section*{Acknowledgement}
We would like to thank Metin Ata, Christian Byrnes, Misao Sasaki, and Puxun Wu for useful discussions. 
This work is supported by the National Key Research and Development Program
of China Grant No. 2021YFC2203004 and No. 2020YFC2201502, by the Key Research Program of the Chinese Academy of
Sciences Grant No. XDPB15, by the CAS Project for Young Scientists in Basic Research YSBR-006,
and by Project 12047503 of the National Natural Science Foundation of China. 
This work is also supported in part by JSPS Grant-in-Aid for Early-Career Scientists No. JP20K14461, 
and by the World Premier International Research Center Initiative (WPI Initiative), MEXT, Japan. 

\appendix

\section{Inflation in Starobinsky model}\label{a:model}

In Appendix \ref{a:model}, we solve the background evolution of the Starobinsky model and derive \eqref{zz} and \eqref{etaa}.

To derive the expression for 
\be\label{def:z}
z=a\mpl\sqrt{2\epsilon}=\frac{a|\dot\phi|}{H},
\ee
we have to solve the background equation of motion. For $\phi>\phi_0$, we assume the inflaton is already well on the slow-roll attractor solution, otherwise additional complexity and large non-Gaussianity may arise. The slow-roll equation of motion for inflaton $\phi$ is 
\be
3 H_{0} \dot{\phi}=-V_{,\phi}=-A_{+}.
\label{CA+}
\ee
For $\phi<\phi_0$, 
\begin{align}
\ddot{\phi}+3 H_{0} \dot{\phi}+A_{-}=&0, \nn
\end{align}
which can be written as
\begin{align}\label{CA-}
\left(\dot{\phi}+\frac{A_{-}}{3 H_{0}}\right)^{\cdot}+3 H_{0}\left(\dot{\phi}+\frac{A_{-}}{3 H_{0}}\right)=0.
\end{align}
The slow-roll solution is simple:
\be
\dot{\phi}+\frac{A_-}{3 H_{0}}=Ce^{-3 H_{0} (t-t_0)}.
\ee
$t_0$ is the moment when $\phi=\phi_0$. Constant $C$ can be determined by equating $\dot\phi$ from \eqref{CA+} and \eqref{CA-} at $t_0$,
\be
C=-\frac{\left( A_+-A_{-}\right)}{3 H_{0}}.
\ee
Besides, we need to transfer to conformal time 
\be\label{def:tau}
\tau=-\int^{t_f}_{t}\frac{\dif t'}{a(t')}=-\frac1{a_0H_0}e^{-H_0(t-t_0)}=\tau_0e^{-H_0(t-t_0)},
\ee
where we have normalized $\tau_0=-1/(a_0H_0)$ when $t=t_0$. $a_0$ and $H_0$  are the scale factor and Hubble parameter at $t_0$, where the latter is given by \eqref{def:H0}. Then from \eqref{def:tau} we have $e^{-H(t-t_0)}=\tau/\tau_0$. Substituting \eqref{CA+} and \eqref{CA-} as well as
\be
a=a_0e^{H_0(t-t_0)}=a_0\left(\frac{\tau_0}{\tau}\right)
\ee
into \eqref{def:z}, we immediately have \eqref{zz}.

By $e$-folding number $N$ counted from $t_0$,
\be
N\equiv\int^t_{t_0}H\dif t\approx H_0(t-t_0),
\ee 
we can write the second slow roll parameter $\eta$ and $z(\tau)$ as
\be
\eta=\frac{1}{\epsilon} \frac{\dif\epsilon}{\dif N}=\frac{\dif }{\dif N} \ln\left(\frac{z^{2}}{2 a^{2}} \right)  =2\left(\frac{1}{z} \frac{\dif z}{\dif N}-1\right),
\label{soleta}
\ee
\be
z=-\frac{a_{0} }{3 H_{0}^{2}}\left( A_- e^{N}+\left(A_{+}-A_{-}\right) e^{-2N} \right).
\label{zinN}
\ee
Substitude (\ref{zinN}) in (\ref{soleta}), one can obtain (\ref{etaa}).

\section{Matching coefficients in Starobinsky model}\label{a:coefficient}

In Appendix \ref{a:coefficient}, we calculate the coefficient $C_i$ and $D_i$ of the solution \eqref{solvk} for the Mukhanov-Sasaki equation, and finally get the coefficients of $\mathcal{O}_{0,1,2}$ in the power spectrum $\mathcal{P_R}$ at horizon crossing given in \eqref{PRk}.

To solve the equation of motion (\ref{R}) for the comoving curvature perturbation $\cal R$, we define the  $v_{k}$,
\be
\mathcal{R}_{k, i}(\tau)=\frac{v_{k}(\tau)}{z_{i}(\tau)}, \quad (i=1,2)
\ee
and get the Mukhanov-Sasaki equation:
\be
v_{k}^{\prime \prime}+\left(k^{2}-\frac{2}{\tau^{2}}\right) v_{k}=0.
\ee
Because of the Wands duality, this equation holds for both the slow-roll and ultra-slow-roll phase, of which the solutions are given by (\ref{solvk}). Asymptoting (\ref{solvk}) to the Bunch-Davies vacuum in the remote past, and then matching the curvature perturbation and its first derivative at $\tau_0$, i.e. $\mathcal{R}_{k, 1}(\tau_0)=\mathcal{R}_{k, 2}(\tau_0)$ and $\mathcal{R}_{k, 1}^{\prime}(\tau_0)=\mathcal{R}_{k, 2}^{\prime}(\tau_0)$, we have 
\begin{align}
&\mathbf{Re}\left(C_{1}(k)\right)=\mathbf{Re}\left(C_{2}(k)\right)=-\frac12\sqrt{\frac{\pi}{\kappa\mathcal{H}_0}}, \\
&\mathbf{Im}\left(C_{2}(k)\right)=\frac{1}{2} \sqrt{\frac{\pi}{\kappa  \mathcal{H}_0}} \cdot  \frac{3}{2}  \left(1-\frac{1}{\Lambda}\right)  \left(\frac{1}{\kappa^{3}}+\frac{1}{\kappa}\right), \\
&\mathbf{Re}\left(D_{2}(k)\right)=-\frac{1}{2} \sqrt{\frac{\pi}{\kappa  \mathcal{H}_0}}  \cdot \frac{3}{2}  \left(1-\frac{1}{\Lambda}\right)  \left[\left(\frac{1}{\kappa^{3}}-\frac{1}{\kappa}\right) \sin (2 \kappa)-\frac{2}{\kappa^{2}} \cos (2 \kappa)\right], \\
&\mathbf{Im}\left(D_{2}(k)\right)=\frac{1}{2} \sqrt{\frac{\pi}{\kappa  \mathcal{H}_0}}  \cdot \frac{3}{2}  \left(1-\frac{1}{\Lambda}\right)  \left[\left(\frac{1}{\kappa^{3}}-\frac{1}{\kappa}\right) \cos (2 \kappa)+\frac{2}{\kappa^{2}} \sin (2 \kappa)\right].
\end{align}
Substitude $\mathcal{R}_{k,i}(\tau)$ into 
\be
\mathcal{P}_{\mathcal{R}}\left(\tau_{k}\right)=\frac{k^{3}}{2 \pi^{2}}\left|\mathcal{R}_{k, i}\left(\tau_{k}\right)\right|^{2}, 
\ee
with $i=1,2$ for $\kappa<\varrho$ and $\kappa\geq\varrho$, respectively, 
we reach the power spectrum (\ref{PRk}), where the coefficients $\mathcal{O}_{0,1,2}$ are
\begin{align}
\mathcal{O}_{0}&=\left(\varrho^{2}+1\right)\left(2 \kappa^{6}+9 \kappa^{4}+18 \kappa^{2}+9\right)\nn \\
&\quad+\Big[ -3 \varrho^{2}\left(7 \kappa^{4}-3\right)-12 \varrho\left(-\kappa^{5}+4 \kappa^{3}+3 \kappa\right)-3\left(3-7 \kappa^{4}\right)\Big]   \cos (2(\varrho-\kappa))\nn \\
&\quad+\Big[  6 \varrho^{2}\left(\kappa^{5}-4 \kappa^{3}-3 \kappa\right)-6 \varrho\left(3-7 \kappa^{4}\right)+6\left(-\kappa^{5}+4 \kappa^{3}+3 \kappa\right) \Big]  \sin (2(\varrho-\kappa)), \\
\mathcal{O}_{1}&=-18\left(\varrho^{2}+1\right)\left(\kappa^{2}+1\right)^{2}\nn \\
&\quad+\Big[  6 \varrho^{2}\left(5 \kappa^{4}-3\right)+12 \varrho\left(-\kappa^{5}+7 \kappa^{3}+6 \kappa\right)+6\left(3-5 \kappa^{4}\right) \Big]  \cos (2(\varrho-\kappa)) \nn \\
&\quad+\Big[ -6 \varrho^{2}\left(\kappa^{5}-7 \kappa^{3}-6 \kappa\right)+6 \varrho\left(6-10 \kappa^{4}\right)-6\left(-\kappa^{5}+7 \kappa^{3}+6 \kappa\right) \Big]   \sin (2(\varrho-\kappa)), \\
\mathcal{O}_{2}&=9\left(\varrho^{2}+1\right)\left(\kappa^{2}+1\right)^{2} \nn \\
&\quad+\Big[ 9 \varrho^{2}\left(1-\kappa^{4}\right)-36 \varrho\left(\kappa^{3}+\kappa\right)+9\left(\kappa^{4}-1\right) \Big]  \cos (2(\varrho-\kappa)) \nn \\
&\quad+\Big[ -18 \varrho^{2}\left(\kappa^{3}+\kappa\right)+18 \varrho\left(\kappa^{4}-1\right)+18\left(\kappa^{3}+\kappa\right) \Big]   \sin (2(\varrho-\kappa)).
\end{align}




\section{Superhorizon enhancement factors}\label{a:alpha}

In Appendix \ref{a:alpha} we calculate the superhorizon enhancement factor $|\alpha_k|$ by the gradient expansion method \cite{Leach:2001zf}, and determine the coefficients in \eqref{alpha}.

It is well known that there are two independent solutions in the equation of motion for the curvature perturbation on superhorizon scales, 
\be
\mathcal{R} \approx A+B \int_{0^-}^{\tau} \frac{\mathrm{d} \tau'}{z^2\left(\tau'\right)}\approx A+\tilde{B} \int_{0^-}^{\tau} \frac{\mathrm{d} \tau'}{a^2\left(\tau'\right)\epsilon(\tau')},
\label{eq:const&decay}
\ee
where $A$, $B$ and $\tilde{B}$ are constants. For a slow-roll inflation, the $A$-term is a constant, while the $B$-term decays rapidly, as $\epsilon$ is approximately invariant. This gives the familiar conclusion that the curvature perturbation $\mathcal{R}$ remains constant after the horizon crossing. However, if the slow-roll condition is violated, the next-to-leading order correction ($\sim\mathcal{O}(k^2)$ in the sense of gradient expansion) to such solutions can be large, thus such arguments are no longer true. 
According to \eqref{eq:const&decay}, this happens when $z^2(\tau) \ll z^2\left(\tau_k\right)$ for $\tau>\tau_k$.

In order to  characterize the behavior of $\mathcal{R}$ after the horizon crossing more accurately, we need to consider the effects from next-order, the so-called gradient expansion method. In this appendix we will briefly review how to calculate the enhancement of the power spectrum by considering the next-to-leading $k^2$-correction to the superhorizon solution \eqref{eq:const&decay}. For higher order corrections, see \cite{Ozsoy:2019lyy}.


We begin with the equation of motion \eqref{R},
\be
\mathcal{R}_k^{\prime \prime}+2 \frac{z^{\prime}}{z} \mathcal{R}_k^{\prime}+k^2 \mathcal{R}_k=0.
\label{eq:Rk}
\ee
We define a set of independent solutions as $u(\tau)$ and $v(\tau)$, assuming $u(\tau)$ as the late-time asymptotic solution at the end of inflation, while $v(\tau)$ decays at that time. Obviously at leading order, $u(\tau)=u_0=A$ is the constant solution in \eqref{eq:const&decay}, while $v(\tau)$ is the other $B$-term ``decaying mode''. Following \cite{Leach:2001zf}, we will call $u$ ``growing mode'' and $v$  ``decaying mode'' in the following discussion. (Note that the terms are different from the main text.) These names are for their asymptotic properties at the end of inflation, but not around the horizon-crossing. Suppose that we already know one of the two linearly independent solutions $u(\tau)$,
the other one, $v(\tau)$, could be solved by the Wronskian 
\be
v(\tau)\propto u(\tau) \int_{0^{-}}^{\tau} \frac{\mathrm{d} \tau'}{z^{2}   u^{2}}.
\label{eq:uAv}
\ee
The normalization coefficient in \eqref{eq:uAv} should be determined at some initial moment, for instance $\tau_k$,
\be
u(\tau_k) =C  v(\tau_k),
\label{eq:C}
\ee
where $C$ is an arbitrary constant which is fixed to be 1 in \cite{Leach:2001zf}.
After matching with \eqref{eq:C}, \eqref{eq:uAv} finally becomes
\be
v(\tau)=\frac{u(\tau)}{C} \frac{D(\tau)}{D\left(\tau_{k}\right)}, \quad D(\tau)\equiv 3 \mathcal{H}_{k} \int_{\tau}^{0^{-}} \mathrm{d} \tau^{\prime} \frac{z^{2}\left(\tau_{k}\right) u^{2}\left(\tau_{k}\right)}{z^{2}\left(\tau^{\prime}\right) u^{2}\left(\tau^{\prime}\right)}.
\label{vinu}
\ee

In general, the curvature perturbation $\mathcal{R}_k$ is a linear combination of the two independent solutions
\be
\mathcal{R}_{k}(\tau)=\alpha_{k}   u(\tau)+C\beta_{k}   v(\tau),
\label{eq:ab}
\ee
where $\alpha_{k}$ and $\beta_{k}$ are constants, and $\alpha_{k}+\beta_{k}=1$ to normalize $\mathcal{R}_k(\tau_k)=u(\tau_k)$ at the horizon crossing.
For slow-roll case, $v(\tau)$ die out rapidly, so we have a scale-invariant curvature perturbation with $\alpha_{k}=1$. However, in the slow-roll violation, the amplitude of the curvature perturbation may get enhanced for $|\alpha_{k}|$,~$|\beta_{k}|\gg 1$. To find the value of $\alpha_{k}$, we take the derivative of \eqref{vinu}, 
\be
v^{\prime}\left(\tau_k\right)=\frac{1}{C}\left(u^{\prime}\left(\tau_k\right)-3 \mathcal{H}_k \frac{u\left(\tau_k\right)}{D_k}\right),
\ee
where $D_k\equiv D(\tau_k)$.
Noting $\alpha_{k}+\beta_{k}=1$, $\mathcal{R}_k$ and $\mathcal{R}'_k$ at $\tau_k$ are given by
\begin{align}\label{eq:ukRk}
\calR_k(\tau_k)&=u(\tau_k),\\
\calR_k'(\tau_k)&=u'(\tau_k)-\frac{3(1-\alpha_k)\calH_ku(\tau_k)}{D_k}.
\end{align}
Combining the above two equations, we can write $\alpha_{k}$ in terms of $\mathcal{R}_k(\tau_k)$ and $\mathcal{R}'_k(\tau_k)$:
\be\label{def:alphak}
\alpha_k=1+\frac{D_k}{3\calH_k}\left(\frac{\calR_k'}{\calR_k}-\frac{u_k'}{u_k}\right)_{\tau=\tau_k}.
\ee

The remaining question is how to calculate $u(\tau)$. At leading order on long-wavelength limit ($k \to 0$), $u(\tau)$ behaves as a constant, $u=u_0$, and 
\be\label{def:v0}
v_0(\tau)=\frac{u_0}{C} \frac{D(\tau)}{D_k},\qquad
D(\tau)\approx 3 \mathcal{H}_k  z^2\left(\tau_k\right) \int_\tau^{0^{-}} \frac{\mathrm{d} \tau^{\prime}}{z^2\left(\tau^{\prime}\right)}.
\ee
The next-leading-order solution of $u(\tau)$ can be obtained by the Green function method
\begin{align}
\label{def:u1}
u(\tau)&=[1+F(\tau)] u_{0},\\
F(\tau) &\approx k^2 \int_{\tau}^{0^{-}} \frac{\mathrm{d} \tau'}{z^2\left(\tau'\right)} \int_{\tau_k}^{\tau'} z^2\left(\tau''\right) \mathrm{d} \tau''.
\label{def:Ftau}
\end{align}
As we commented, $F(\tau)$  becomes large when $z^{2}(\tau) \ll z^{2}\left(\tau_{k}\right)$, which can no longer be omitted. However, as we can see from \eqref{def:Ftau}, $u(\tau)$ now depends on time. Suppose $z^{2}(\tau) \ll z^{2}\left(\tau_{k}\right)$ when $\tau_k < \tau < \tau_*$, $F(\tau)$ becomes large until $\tau_*$, and then it decays as usual. This violates the definition of $u(\tau)$, which should converges a constant at infinite future. To solve this problem, considering that $F(\tau)$ behaves similarly to that of the lowest order solutions of $v(\tau)$, we subtract this time-dependent part from \eqref{def:u1}, and define
\be
u=\big[1+F(\tau)\big]u_0-CF_kv_0(\tau),
\ee
where $F_k=F(\tau_k)$. Now the new growing mode is a constant on superhorizon scales, $u\approx u_0$, or at least $u(0)\approx u(\tau_k)$. At the horizon crossing, we have
\begin{align}
u\left(\tau_k\right) &=\left[1+F_k\right] u_0-F_k u_0=u_0=Cv_0\left(\tau_k\right), \\
u'\left(\tau_k\right) &=F'\left(\tau_k\right) u_0-CF_k v'_0\left(\tau_k\right)=CF'\left(\tau_k\right) v_0\left(\tau_k\right)-CF_k v'_0\left(\tau_k\right),
\end{align}
thus
\be\label{u'/u}
\left[\frac{u^{\prime}}{u}\right]_{\tau=\tau_k}=F^{\prime}\left(\tau_k\right)-F_k \left[\frac{v_0^{\prime}}{v_0}\right]_{\tau=\tau_k}=-F_k \left[\frac{v_0^{\prime}}{v_0}\right]_{\tau=\tau_k}=\frac{3\calH_k}{D_k}F_k.
\ee
In the second step we use $F'(\tau_k)=0$ as the integral of $F_k(\tau)$ is from $\tau_k$ to $\tau$, while in the last step we use the definition \eqref{def:v0}. Therefore substituting \eqref{u'/u} into \eqref{def:alphak}, we immediately have
\be\label{alphak}
\alpha_k=1+\frac{\calR_k'(\tau_k)}{3\calH_k\calR_k(\tau_k)}D_k-F_k.
\ee
Now, the curvature perturbation at the end of inflation can be simply achieved by taking the $\tau\to0^-$ limit of \eqref{eq:ab}:
\be
\calR(0^-)=\alpha_ku(0^-)\approx\alpha_ku(\tau_k)=\alpha_k\calR_k(\tau_k),
\ee
where the first equality is because $v(\tau)$ dies out rapidly when approaching the end of inflation.

Applying the aforementioned formulae to the Starobinsky model, we have the expression for the power spectrum, $\calP_\calR(0^-)=|\alpha_k|^2\calP_\calR(\tau_k)$, where $\alpha_k$ is given \eqref{alphak}. Depending on whether the wavelength of a mode exceeds horizon scale during the first stage ($\tau_{k}<\tau_{0}$) or second stage ($\tau_{k}>\tau_{0}$), the integral in \eqref{Ddir} and \eqref{Fdir} should be performed in different segments: for $\tau_{k}<\tau_{0}$, $\kappa<\varrho$:
\begin{align}
D_k=&3 \frac{k}{\varrho}   z_{1}^{2}\left(\tau_{k}\right)\left\{\int_{\tau_{k}}^{\tau_{0}} \frac{\mathrm{d} \tau^{\prime}}{z_{1}^{2}\left(\tau^{\prime}\right)}+\int_{\tau_{0}}^{0^{-}} \frac{\mathrm{d} \tau^{\prime}}{z_{2}^{2}\left(\tau^{\prime}\right)}\right\},\\
F_k=&k^{2} \int_{\tau_{k}}^{\tau_{0}} \frac{\mathrm{d} \tau^{\prime}}{z_{1}^{2}\left(\tau^{\prime}\right)} \int_{\tau_{k}}^{\tau^{\prime}} z_{1}^{2}\left(\tau^{\prime \prime}\right) \mathrm{d} \tau^{\prime \prime} \nn\\
+&k^{2} \int_{\tau_{0}}^{0^{-}} \frac{\mathrm{d} \tau^{\prime}}{z_{2}^{2}\left(\tau^{\prime}\right)}\left\{\int_{\tau_{k}}^{\tau_{0}} z_{1}^{2}\left(\tau^{\prime \prime}\right) \mathrm{d} \tau^{\prime \prime}+\int_{\tau_{0}}^{\tau^{\prime}} z_{2}^{2}\left(\tau^{\prime \prime}\right) \mathrm{d} \tau^{\prime \prime}\right\}.
\end{align}
For $\tau_{k}\geq\tau_{0}$, $\kappa\geq\varrho$:
\begin{align}
D_k=&3 \frac{k}{\varrho}   z_{2}^{2}\left(\tau_{k}\right) \int_{\tau_{k}}^{0^{-}} \frac{\mathrm{d} \tau^{\prime}}{z_{2}^{2}\left(\tau^{\prime}\right)},\\
F_k=&k^{2} \int_{\tau_{k}}^{0^{-}} \frac{\mathrm{d} \tau^{\prime}}{z_{2}^{2}\left(\tau^{\prime}\right)} \int_{\tau_{k}}^{\tau^{\prime}} z_{2}^{2}\left(\tau^{\prime \prime}\right) \mathrm{d} \tau^{\prime \prime}.
\end{align}
Here we have defined $z_{i}(\tau)=z(\tau)$, $i=1,2$ representing before and after $\tau_0$, respectively.

The dependence of the integral $D_k$ and $F_k$ on $\kappa$ is given by \eqref{Ddir} and \eqref{Fdir}. Substituding these results to \eqref{def:alpha}, together with the curvature perturbation $\calR_k$ and $\calR_k'$ at the horizon crossing, we see that $\left|\alpha_k\right|^2$ can be expressed by \eqref{alpha}, with the following coefficients.
\begin{align}
&s_{0}=\frac{1}{1+\varrho^{2}}\left[\frac{\kappa^{6}}{9}-\frac{4 \varrho}{15} \kappa^{5}+\frac{\varrho\left(\varrho^{2}-6\right)}{9} \kappa^{3}-\frac{2\left(\varrho^{4}-3 \varrho^{2}-6\right)}{15} \kappa^{2}+\frac{\varrho^{6}-3 \varrho^{4}+36}{36}\right]+\frac{4 \kappa^{4}}{25}, \\
&s_{1}=\frac{1}{1+\varrho^{2}}\left[-\frac{2}{9} \kappa^{6}+\frac{8 \varrho}{15} \kappa^{5}-\frac{\varrho\left(\varrho^{2}-6\right)}{9} \kappa^{3}+\frac{2\left(\varrho^{4}-3 \varrho^{2}-6\right)}{15} \kappa^{2}\right]-\frac{8 \kappa^{4}}{25}, \\
&s_{2}=\frac{1}{1+\varrho^{2}}\left[\frac{\kappa^{6}}{9}-\frac{4 \varrho}{15} \kappa^{5}\right]+\frac{36 \kappa^{4}}{225}.
\end{align}
\begin{align}
W_{1}=&\left(2 \varrho^{5}  \left(1-\Lambda\right)+5\left(6 -\varrho^{2}\right) \kappa^{3}\right)^{2},\\
W_{2}=&\left(2 \varrho^{5}  \left(1-\Lambda\right)+5\left(6 -\varrho^{2}\right) \kappa^{3}\right)   10,\\
W_{3}=&64   \kappa^{12} \varrho^6 \left(\left(1-\Lambda\right)-\left(\kappa/\varrho\right)^{3}\right)^{2}\Lambda^{4},\\
W_{0}=&W_{4}^{2}+W_{5}^{2},\\
\mathcal{W}=&W_{5}  \left(W_{9}-W_{6}\right)+W_{4}  \left(W_{7}-W_{8}\right).
\end{align}
\begin{align}
W_{4}=&G_{1}  \left(1-\Lambda\right)+\sqrt{\frac{1}{\varrho}} \left[2 \kappa^{3}   \left(\frac{\sin (\varrho)}{\varrho}-\cos (\varrho)\right)\right]  \Lambda,
\end{align}
\begin{align}
W_{5}=&G_{2}  \left(1-\Lambda \right)+\sqrt{\frac{1}{\varrho}} \left[-2 \kappa^{3}   \left(\frac{\cos (\varrho)}{\varrho}+\sin (\varrho)\right)\right]  \Lambda.
\end{align}
\begin{align}\nn
W_{6}=&\left(G_{3}  \left(1-\Lambda\right)-2 \kappa^{3}  \Lambda\right) \nn \\
 &\left\{\sqrt{\frac{1}{\varrho}} \left[\left(\varrho \kappa^{3}-\varrho^{4}\right)   \left(\frac{3 \cos (\varrho)}{\varrho^{2}}+\frac{3 \sin (\varrho)}{\varrho}-2 \cos (\varrho)\right)-\left(3 \varrho^{3}+3 \kappa^{3}\right)   \left(\frac{\cos (\varrho)}{\varrho}+\sin (\varrho)\right)\right]\right.\nn \\
+&\left. \sqrt{\frac{1}{\varrho}} \left[\varrho^{4}   \left(\frac{3 \cos (\varrho)}{\varrho^{2}}+\frac{3 \sin (\varrho)}{\varrho}-2 \cos (\varrho)\right)+3 \varrho^{3}   \left(\frac{\cos (\varrho)}{\varrho}+\sin (\varrho)\right)\right]    \Lambda\right\},
\end{align}
\begin{align}\nn
W_{7}=&\left(G_{3}  \left(1-\Lambda\right)+2 \kappa^{3}   \Lambda\right) \nn \\
 &\left\{\sqrt{\frac{1}{\varrho}} \left[-\left(\varrho \kappa^{3}-\varrho^{4}\right)   \left(\frac{3 \sin (\varrho)}{\varrho^{2}}-\frac{3 \cos (\varrho)}{\varrho}-2 \sin (\varrho)\right)+\left(3 \varrho^{3}+3 \kappa^{3}\right)   \left(\frac{\sin (\varrho)}{\varrho}-\cos (\varrho)\right)\right] \right.\nn \\
+&\left. \sqrt{\frac{1}{\varrho}} \left[-\varrho^{4}   \left(\frac{3 \sin (\varrho)}{\varrho^{2}}-\frac{3 \cos (\varrho)}{\varrho}-2 \sin (\varrho)\right)-3 \varrho^{3}   \left(\frac{\sin (\varrho)}{\varrho}-\cos (\varrho)\right)\right]   \Lambda \right\},
\end{align}
\begin{align}\nn
W_{8}=&\left(3\left(1+\kappa^{2}\right)-G_{4}\right)  \left(1-\Lambda\right) \nn \\
 &\left\{\sqrt{\frac{1}{\varrho}} \left[\left(\varrho \kappa^{3}-\varrho^{4}\right)   \left(\frac{3 \cos (\varrho)}{\varrho^{2}}+\frac{3 \sin (\varrho)}{\varrho}-2 \cos (\varrho)\right)-\left(3 \varrho^{3}+3 \kappa^{3}\right)   \left(\frac{\cos (\varrho)}{\varrho}+\sin (\varrho)\right)\right]\right.\nn \\
+&\left. \sqrt{\frac{1}{\varrho}} \left[\varrho^{4}   \left(\frac{3 \cos (\varrho)}{\varrho^{2}}+\frac{3 \sin (\varrho)}{\varrho}-2 \cos (\varrho)\right)+3 \varrho^{3}   \left(\frac{\cos (\varrho)}{\varrho}+\sin (\varrho)\right)\right]    \Lambda\right\},
\end{align}
\begin{align}\nn
W_{9}=&\left(3\left(1+\kappa^{2}\right)+G_{4}\right)  \left(1-\Lambda\right)\nn \\
 &\left\{\sqrt{\frac{1}{\varrho}} \left[-\left(\varrho \kappa^{3}-\varrho^{4}\right)   \left(\frac{3 \sin (\varrho)}{\varrho^{2}}-\frac{3 \cos (\varrho)}{\varrho}-2 \sin (\varrho)\right)+\left(3 \varrho^{3}+3 \kappa^{3}\right)   \left(\frac{\sin (\varrho)}{\varrho}-\cos (\varrho)\right)\right] \right.\nn \\
+&\left. \sqrt{\frac{1}{\varrho}} \left[-\varrho^{4}   \left(\frac{3 \sin (\varrho)}{\varrho^{2}}-\frac{3 \cos (\varrho)}{\varrho}-2 \sin (\varrho)\right)-3 \varrho^{3}   \left(\frac{\sin (\varrho)}{\varrho}-\cos (\varrho)\right)\right]   \Lambda \right\}.
&\nn 
\end{align}
\begin{align}
G_{1}=&\sqrt{\frac{1}{\varrho}} \left[3\left(\kappa^{2}+1\right)   \left(\frac{\cos (\varrho)}{\varrho}+\sin (\varrho)\right)\right]\nn \\
+&\sqrt{\frac{1}{\varrho}} \left[6 \kappa   \left(\frac{\sin (\varrho)}{\varrho}-\cos (\varrho)\right)+3\left(\kappa^{2}-1\right)   \left(\frac{\cos (\varrho)}{\varrho}+\sin (\varrho)\right)\right]   \cos (2 \kappa)\nn \\
+&\sqrt{\frac{1}{\varrho}} \left[-6 \kappa   \left(\frac{\cos (\varrho)}{\varrho}+\sin (\varrho)\right)+3\left(\kappa^{2}-1\right)   \left(\frac{\sin (\varrho)}{\varrho}-\cos (\varrho)\right)\right]   \sin (2 \kappa), 
\end{align}
\begin{align}
G_{2}=&\sqrt{\frac{1}{\varrho}} \left[3\left(\kappa^{2}+1\right)   \left(\frac{\sin (\varrho)}{\varrho}-\cos (\varrho)\right)\right]\nn \\
-&\sqrt{\frac{1}{\varrho}} \left[-6 \kappa   \left(\frac{\cos (\varrho)}{\varrho}+\sin (\varrho)\right)+3\left(\kappa^{2}-1\right)   \left(\frac{\sin (\varrho)}{\varrho}-\cos (\varrho)\right)\right]   \cos (2 \kappa)\nn \\
+&\sqrt{\frac{1}{\varrho}} \left[6 \kappa   \left(\frac{\sin (\varrho)}{\varrho}-\cos (\varrho)\right)+3\left(\kappa^{2}-1\right)   \left(\frac{\cos (\varrho)}{\varrho}+\sin (\varrho)\right)\right]   \sin (2 \kappa), 
\end{align}
\begin{align}
G_{3}=&6 \kappa   \cos (2 \kappa)-3\left(1-\kappa^{2}\right)   \sin (2 \kappa), 
\end{align}
\begin{align}
G_{4}=&3\left(1-\kappa^{2}\right)   \cos (2 \kappa)+6 \kappa   \sin (2 \kappa).
\end{align}

\section{Coefficients in constant-roll parameterization}\label{a:constantroll}

In Appendix \ref{a:constantroll} we list the coefficients $\mathcal{G}_{0,1,2,3,4}$ in the power spectrum \eqref{PR3} of the constant-roll model \eqref{stepfunc}.

The solution of curvature perturbation at the final slow-roll stage is already given by Ref.\cite{Byrnes:2018txb}
\be
\mathcal{R}_{k}(\tau)=i \frac{H_{0}}{M_{\mathrm{pl}}} \frac{1}{\sqrt{4k^{3}  \epsilon_{0}  e^{-6N_\mathrm{USR}} }}\left[c(\kappa)(1+i k \tau) e^{-i k \tau}-d(\kappa)(1-i k \tau) e^{i k \tau}\right],
\ee
where the coefficients of two modes are
\begin{align}
&c\left(\kappa\right)=-\frac{1}{4   \kappa^{3} \kappa_{1}^{3}}\left\{9 e^{2 i\left(\kappa-\kappa_{1}\right)}\left(\kappa+i\right)^{2}\left(\kappa_{1}-i\right)^{2}-\left(\kappa^{2}\left(2 \kappa-3 i\right)-3 i\right)\left(\kappa_{1}^{2}\left(2 \kappa_{1}+3 i\right)+3 i\right)\right\}, \\
&d\left(\kappa\right)=\frac{e^{2 i\left(\kappa+\kappa_{1}\right)}}{4   \kappa^{3} \kappa_{1}^{3}}\left\{3 e^{-2 i \kappa_{1}}\left(3+\kappa_{1}^{2}\left(3+2 i \kappa_{1}\right)\right)\left(\kappa+i\right)^{2}-3 i e^{-2 i \kappa}\left(\kappa^{2}\left(2 \kappa-3 i\right)-3 i\right)\left(\kappa_{1}+i\right)^{2}\right\},
\end{align}
with $\kappa\equiv -k\tau_0$ and $\kappa_1\equiv-k\tau_1$. This is the final solution after matching two boundary conditions at $\tau_0$ and $\tau_1$, so the power spectrum is given by
\be
\mathcal{P}_{\mathcal{R}}(\kappa)=\lim _{\tau \to 0^{-}} \frac{k^{3}}{2 \pi^{2}}\left|\mathcal{R}_{k}(\tau)\right|^{2}. \nn 
\ee
We split the components of different frequencies in constant-roll model (\ref{PR3}), and the dependence of the coefficients in front of each component on $\kappa$ is as follows:
\begin{align}
&\mathcal{G}_{0}(\kappa)=4\left(9+18 \kappa^{2}+9 \kappa^{4}+2 \kappa^{6}\right)\left(9+18 \kappa_{1}^{2}+9 \kappa_{1}^{4}+2 \kappa_{1}^{6}\right),\\
&\mathcal{G}_{1}(\kappa)=3\left(i+\kappa\right)^{2}\left(3+3 \kappa^{2}-2 i  \kappa^{3}\right)\left(3 i+3 i  \kappa_{1}^{2}-2 \kappa_{1}^{3}\right)^{2}, \\
&\mathcal{G}_{2}(\kappa)=18\left(i+\kappa\right)^{2}\left(3 i+3 i  \kappa^{2}+2 \kappa^{3}\right)\left(-i+\kappa_{1}\right)^{2}\left(3i+3 i  \kappa_{1}^{2}-2 \kappa_{1}^{3}\right), \\
&\mathcal{G}_{3}(\kappa)=27 i\left(i+\kappa\right)^{2}\left(3 i+3 i \kappa^{2}+2 \kappa^{3}\right)\left(-i+\kappa_{1}\right)^{4}, \\
&\mathcal{G}_{4}(\kappa)=6\left(i+\kappa_{1}\right)^{2}\left(9+18 \kappa^{2}+9 \kappa^{4}+2 \kappa^{6}\right)\left(3+3 \kappa_{1}^{2}+2 i \kappa_{1}^{3}\right). 
\end{align}
%

\bibliography{ref}

\end{document}